\documentclass[iop,apj,numberedappendix,twocolappendix]{emulateapj}
\usepackage{geometry} 
\usepackage{natbib}
\usepackage{graphicx}
\usepackage{amssymb}
\usepackage{amsmath}
\usepackage{epsfig}
\usepackage{epstopdf}
\usepackage[usenames,dvipsnames]{color}
\usepackage[normalem]{ulem}
\usepackage{sidecap}
\usepackage{placeins}

\newcommand{\sst}{\textsuperscript{st}\ }
\newcommand{\snd}{\textsuperscript{nd}\ }
\newcommand{\srd}{\textsuperscript{rd}\ }
\newcommand{\sth}{\textsuperscript{th}\ }
\newcommand{\beqn}{\begin{equation}}
\newcommand{\eeqn}{\end{equation}}

\newcommand{\ie}{\emph{i.e.,}\ }
\newcommand{\cf}{\emph{cf.}\ }
\newcommand{\eg}{\emph{e.g.,}\ }

\newcommand{\kms}{km~s$^{-1}$}

\newcommand{\vlsr}{$v_{_\mathrm{LSR}}$}
\newcommand{\dtan}{$d_\mathrm{tan}$}
\newcommand{\dml}{$d_{_\mathrm{ML}}$}
\newcommand{\dbar}{$\overline{d}$}
\newcommand{\dsun}{$d_{_\sun}$}
\newcommand{\rgal}{$R_\mathrm{gal}$}
\newcommand{\spitzer}{\emph{Spitzer}}

\newcommand{\cc}{cm$^{-3}$}

\newcommand{\neff}{$n_\mathrm{eff}$}
\newcommand{\pchoose}{$P_{_\mathrm{ML}}$}

\newcommand{\HI}{\ion{H}{1}}
\newcommand{\HII}{\ion{H}{2}}
\newcommand{\htwo}{H$_2$}
\newcommand{\hcop}{HCO$^{^+}$}
\newcommand{\nnhp}{N$_2$H$^{^+}$}
\newcommand{\nhhh}{NH$_3$}
\newcommand{\thco}{$^{13}$CO}
\newcommand{\twco}{$^{12}$CO}

\newcommand{\water}{H$_2$O}
\newcommand{\methanol}{CH$_3$OH}

\newcommand{\lbd}{($\ell,b$,\dsun)}

\newcommand{\omni}{\texttt{distance-omnibus}}
\newcommand{\tsource}{$T_\mathrm{source}(v)$}
\newcommand\pasa{\ref@jnl{PASA}}

\shorttitle{BGPS XII.  Distance Catalog}
\shortauthors{Ellsworth-Bowers \emph{et al.}}

\slugcomment{Accepted Version: \today}

\begin{document}

\bibliographystyle{apj}

\title{The Bolocam Galactic Plane Survey. XII. Distance Catalog Expansion Using Kinematic Isolation of Dense Molecular Cloud Structures With \thco(1-0)}

\author{Timothy P. Ellsworth-Bowers\altaffilmark{1,2}, Erik Rosolowsky\altaffilmark{3}, Jason Glenn\altaffilmark{1}, Adam Ginsburg\altaffilmark{4}, Neal J. Evans II\altaffilmark{5}, Cara Battersby\altaffilmark{6}, Yancy L. Shirley\altaffilmark{7,8}, Brian Svoboda\altaffilmark{7}}
\altaffiltext{1}{CASA, University of Colorado, UCB 389, Boulder, CO 80309, USA}
\altaffiltext{2}{email: \texttt{timothy.ellsworthbowers@colorado.edu}}
\altaffiltext{3}{Department of Physics, 4-183 CCIS, University of Alberta, Edmonton, AB T6G 2E1, Canada}
\altaffiltext{4}{European Southern Observatory,
Karl-Schwarzschild-Stra\ss e 2, 85748,
Garching bei M\"{u}nchen, Germany}
\altaffiltext{5}{Department of Astronomy, University of Texas, 2515 Speedway, Stop C1400, Austin, TX 78712, USA}
\altaffiltext{6}{Harvard-Smithsonian Center for Astrophysics, 60 Garden Street, Cambridge, MA 02138 USA}
\altaffiltext{7}{Steward Observatory, University of Arizona, 933 North Cherry Avenue, Tucson, AZ 85721, USA}
\altaffiltext{8}{Adjunct Astronomer at the National Radio Astronomy Observatory}

\begin{abstract}

We present an expanded distance catalog for 1,710 molecular cloud structures identified in the Bolocam Galactic Plane Survey (BGPS) version 2, representing a nearly threefold increase over the previous BGPS distance catalog.  We additionally present a new method for incorporating extant data sets into our Bayesian distance probability density function (DPDF) methodology.  To augment the dense-gas tracers (\eg \hcop(3-2), \nhhh(1,1)) used to derive line-of-sight velocities for kinematic distances, we utilize the Galactic Ring Survey \thco(1-0) data to morphologically extract velocities for BGPS sources.  The outline of a BGPS source is used to select a region of the GRS \thco\ data, along with a reference region to subtract enveloping diffuse emission, to produce a line profile of \thco\ matched to the BGPS source.  For objects with a \hcop(3-2) velocity, $\approx95\%$ of the new \thco(1-0) velocities agree with that of the dense gas.  A new prior DPDF for kinematic distance ambiguity (KDA) resolution, based on a validated formalism for associating molecular cloud structures with known objects from the literature, is presented.  We demonstrate this prior using catalogs of masers with trigonometric parallaxes and \HII\ regions with robust KDA resolutions.  The distance catalog presented here contains well-constrained distance estimates for 20\% of BGPS V2 sources, with typical distance uncertainties $\lesssim 0.5$~kpc.  Approximately 75\% of the well-constrained sources lie within 6~kpc of the Sun, concentrated in the Scutum-Centarus arm.  Galactocentric positions of objects additionally trace out portions of the Sagittarius, Perseus, and Outer arms in the first and second Galactic quadrants, and we also find evidence for significant regions of interarm dense gas.

\end{abstract}



\section{INTRODUCTION}\label{ch3:intro}

Continuum surveys of the Galactic plane at (sub-)millimeter wavelengths (BGPS, \citealp{Aguirre:2011}, \citealp{Ginsburg:2013}; ATLASGAL, \citealp{Schuller:2009}; Hi-GAL, \citealp{Molinari:2010a}), as well as all-sky cosmic microwave background missions \citep[\eg][]{Planck:er22,Planck:er23,Planck:ir23}, have cataloged tens of thousands of dense molecular cloud cores and clumps; the possible precursors to stellar clusters, OB associations, or smaller stellar groups.  Derived distances to and physical properties of these objects may answer several outstanding questions about massive star formation \citep{Kennicutt:2012}.  What is the Galactic distribution of massive star formation in the Milky Way?  What is the clump mass function and its relationship to the stellar initial mass function?  What are the evolutionary processes of the dense interstellar medium?

While stellar and extragalactic studies may make use of standard(izable) candles, no intrinsic luminosity-distance relationship exists for molecular cloud structures.  Distance estimates for these objects must, therefore, rely upon ancillary data.  The distance probability density function (DPDF) formalism introduced by \citet[][hereafter EB13]{EllsworthBowers:2013} provides a means for combining an arbitrary number of ancillary data sets to derive distance estimates for molecular cloud structures detected by continuum surveys.  The primary input to this Bayesian method is a kinematic distance, whereby a measured line-of-sight velocity (\vlsr) is combined with a Galactic rotation curve.  Geometric considerations reveal that two heliocentric distances (\dsun) may correspond to a single \vlsr\ for Galactocentric locations within the solar circle, the so-called kinematic distance ambiguity (KDA).  The DPDF formalism utilizes data from other wavelengths to place priors on the kinematic distance likelihood in an effort to resolve the KDA.  This paper expands on the DPDF methodology by introducing a new source for deriving kinematic distance likelihoods and also a validated formalism for the association of detected molecular cloud structures with published catalogs of objects having robust distance measurements.

Other than being hindered by the KDA, the use of kinematic distances is limited only by the particulars of the chosen rotation curve and molecular line detection rates.  Differences between rotation curves in the Galactic orbital velocity and the distance to the Galactic center can lead to $\gtrsim 20\%$ differences in distances, or $\gtrsim 40\%$ uncertainties in mass \citep[][]{Anderson:2012}.  In terms of molecular line detection, the low-$J$ transitions of \twco, while ubiquitous throughout the Galactic plane, are excited at low density, are generally optically thick, and do not provide a unique tracer for the dense molecular cloud structures detected in continuum surveys around $\lambda = 1$~mm.  Other molecular transitions excited only at higher density \citep[\eg \hcop(3-2), \nhhh(1,1), CS(2-1);][]{Evans:1999} have been targeted by pointed observations in an effort to derive single-component \vlsr\ for these structures \citep[\cf][]{Shirley:2013,Wienen:2012,Jackson:2008}.  Detection rates using these transitions, however, can be somewhat poor ($\sim 50\% - 85\%$), refocusing attention on CO for measuring \vlsr.

The largest single Galactic molecular line data set to date is the compiled \twco(1-0) surveys presented in \citet[][]{Dame:2001}.  With a resolution of several arcminutes, this mosaic revealed the basic structure of the Milky Way as seen in giant molecular clouds (GMCs).  The combination of opacity with the low resolution means that structure visible in the \citeauthor{Dame:2001} data represents primarily the surfaces of GMCs, and not the denser interior clumps.  The isotopologue \thco\ is $\sim50-90\times$ optically thinner (depending on Galactocentric location) and is a better tracer of structure within GMCs.  In large-scale \thco(1-0) data, such as the Galactic Ring Survey \citep[GRS;][]{Jackson:2006}, it is simplest to identify and study GMC-scale structures \citep[][]{RomanDuval:2009,RomanDuval:2010}, although it is possible to identify the clump-scale structures usually detected by continuum surveys \citep{Rathborne:2009}.  Because molecular cloud clumps exist within an envelope of more diffuse gas, disentangling the emission from a single continuum-detected object is challenging \citep[][]{Simon:2006b,Battisti:2014}.  The low effective density for \thco(1-0) excitation also significantly increases the detection of multiple cloud structures along a given line of sight.  Direct \vlsr\ comparison of the brightest \thco\ peak along the line of sight towards continuum-detected objects with the \vlsr\ from a dense-gas tracer like \hcop(3-2) yields a $\sim81\%$ agreement rate \citep{Shirley:2013}.  In this work we present a technique for extracting a \thco\ \vlsr\ for molecular cloud structures using the morphology and flux density from continuum data as prior information to increase the agreement rate to $\sim95\%$.

The kinematic distance likelihoods from \thco\ data expand the catalog of objects with detected \vlsr\ beyond the sample of dense-gas surveys, but the longitude and velocity range of the GRS implies all new sources will be subject to the KDA.  To resolve the ambiguities, EB13 presented prior DPDFs based on eight-micron absorption features (EMAFs) and a simple model for the Galactic distribution of molecular gas; we expand upon that set here.  With the recent publication of distance catalogs for large numbers of objects associated with sites of high-mass star formation it becomes economical to construct a mechanism for associating these distances with continuum-identified molecular cloud structures based on angular and velocity separations.  We demonstrate this new prior with ``gold-standard'' trigonometric parallax measurements to \methanol\ and \water\ masers from the Bar and Spiral Structure Legacy \citep[BeSSeL;][]{Brunthaler:2011} Survey and Japanese VERA project\footnote{\texttt{http://veraserver.mtk.nao.ac.jp}} as well as robust KDA resolutions to \HII\ regions using \HI\ absorption techniques from the \HII\ Region Discovery Surveys \citep[HRDS;][]{Bania:2010,Bania:2012}.

The recent re-reduction and release of the BGPS version 2 data and associated catalog \citep[][hereafter G13]{Ginsburg:2013} provides a solid basis for derivation of posterior DPDFs for molecular cloud structures across the Galactic plane.  We present a new distance catalog based on the BGPS V2.1 catalog, the methods presented in EB13, and those developed here.  While molecular cloud clumps are the objects of interest for understanding the larger process of massive star formation \citep[][]{McKee:2007}, surveys like the BGPS are sensitive to scales ranging from cores that will from a single stellar system to GMCs themselves at the largest distances.  We therefore use the term ``molecular cloud structure'' when discussing detected objects to account for this uncertainty.  A forthcoming work will further discuss the nature of BGPS sources in terms of physical properties derived using DPDF distance estimates.

This paper is organized as follows.  Section~\ref{ch3:data} describes the data used and the distance-computation code.  An overview of the DPDF formalism and relevant Galactic kinematics is given in Section~\ref{ch3:dpdf}.  The estimation of molecular cloud clump \vlsr\ from \thco\ data is described in Section~\ref{ch3:method}, while the new prior DPDFs are outlined in Section~\ref{ch3:priors}.  Section~\ref{ch3:results} presents the results, and implications of this work are discussed in Section~\ref{ch3:discuss}.   A summary is presented in Section~\ref{ch3:summary}.


\section{DATA}\label{ch3:data}

\subsection{The Bolocam Galactic Plane Survey}\label{data:bgps}

The Bolocam Galactic Plane Survey version 2 \citep[BGPS;][G13]{Aguirre:2011}, is a $\lambda = 1.1$~mm continuum survey covering 192~deg$^2$ at 33\arcsec\ resolution.  It is one of the first large-scale blind surveys of the Galactic plane in this region of the spectrum, covering $-10\degr \leq \ell \leq 90\degr$ with at least $|b| \leq 0\fdg5$, plus selected regions in the outer Galaxy.  For a map of BGPS V2.0 coverage and details about observation methods and the data reduction pipeline, see G13.  From the BGPS V2.0 images, 8,594 millimeter dust-continuum sources were identified using a custom extraction pipeline.  BGPS V2.0 pipeline products, including image mosaics and the catalog, are publicly available.\footnote{Available through IPAC at\\ \texttt{http://irsa.ipac.caltech.edu/data/BOLOCAM\_GPS}}

The BGPS data pipeline removes atmospheric signal using a principle component analysis technique that discards common-mode time-stream signals correlated among bolometers in the focal plane array.  The pipeline attempts to iteratively identify astrophysical signal and prevent that signal from being discarded with the atmospheric signal.  This filtering may be characterized by an angular transfer function that passes scales from approximately 33\arcsec\ to 6\arcmin\ (see G13 for a full discussion).  The effective angular size range of detected BGPS sources therefore corresponds to anything from molecular cloud cores up to entire GMCs depending on heliocentric distance \citep{Dunham:2011c}.

\subsection{Molecular Line Spectroscopic Observations}\label{data:spec}

Several spectroscopic follow-up programs have been conducted to observe BGPS version 1 sources in molecular transition lines with excitation effective density\footnote{The effective density required to produce a line with a main beam brightness temperature of 1~K; may be up to several orders of magnitude smaller than the critical density \citep{Evans:1999}.} \neff~$\gtrsim 10^3$~\cc.  All 6,194 BGPS V1 sources with $\ell \geq 7\fdg5$ were simultaneously observed in the $J$=3$-$2 rotational transitions of \hcop\ ($\nu = 267.6$~GHz) and \nnhp\ ($\nu = 279.5$~GHz) with the Heinrich Hertz Submillimeter Telescope (HHT) on Mt.~Graham, Arizona \citep[][hereafter S13]{Schlingman:2011,Shirley:2013}.  Detectability in this \hcop\ line is a strong function of millimeter flux density, and only $\approx 50$\% of pointings produced a $\geq3\sigma$ detection.  Detected sources, however, rarely (1\%) have multiple velocity components, yielding a very robust kinematic catalog.

To complement the HHT survey, 555 BGPS V1 sources in the range  $29\degr \leq \ell \leq 31\degr$ were observed in the $J$=2$-$1 rotational transition of CS ($\nu = 97.98$~GHz) using the Arizona Radio Observatory 12m telescope on Kitt Peak (Y. Shirley, 2012, private communication).  The lower \neff\ of this transition led to CS(2-1) detections for 45\% of sources not detected by the HHT survey in this region.  A further study of this Galactic longitude range was observed in November 2013 at the HHT using the $J$=2$-$1 rotational transition of C$^{18}$O ($\nu = 219.6$~GHz).  From these observations, 190 unique detections were extracted (M. Lichtenberger, 2014, private communication).  

Finally, a total of 686 BGPS V1 sources in the inner Galaxy \citep{Dunham:2011c} and Gemini OB1 molecular cloud complex \citep{Dunham:2010} were observed in the lowest inversion transition lines of \nhhh\ near 24 GHz with the Robert F. Byrd Green Bank Telescope.  The (1,1) inversion is the strongest \nhhh\ transition at the cold temperatures of BGPS sources ($T \approx 20$~K), and is the sole ammonia inversion transition used for velocity fitting.  Detection rates for \nhhh(1,1) range from 61\% in the Gemini OB1 molecular cloud to 72\% in the inner Galaxy.

\subsection{The Galactic Ring Survey}\label{data:grs}

The Galactic Ring Survey \citep[GRS;][]{Jackson:2006} conducted large-scale observations of \thco(1-0) emission in the northern Galactic plane using the FCRAO 14-m telescope.  The survey covers the longitude range $18\degr \leq \ell \leq 55\fdg7$, with select regions down to $\ell \approx 14\degr$, and spans $|b| \leq 1\degr$, with angular resolution and sampling of 46\arcsec\ and 22\arcsec, respectively.  Velocity coverage is $-5$~\kms~$\leq$ \vlsr~$\leq 135$~\kms\ for $\ell \leq 40\degr$, and $-5$~\kms~$\leq$ \vlsr~$\leq 85$~\kms\ for $\ell \geq 40\degr$, with spectral resolution of 0.212~\kms.  Because the GRS aimed to study the Galactic molecular ring at $R_\mathrm{gal} \approx 4-5$~kpc, negative \vlsr\ in this region were omitted.  The GRS data have an rms sensitivity of 0.13~K.

\subsection{The \omni\ Code Base}\label{data:d-o}

The code used to compute DPDFs for this work, while developed for use with the BGPS, is easily configurable for use with other large (sub-)millimeter surveys.  It is written in the Interactive Data Language (\texttt{IDL}),\footnote{\texttt{http://www.exelisvis.com/ProductsServices/IDL.aspx}} and is publicly available.\footnote{\texttt{https://github.com/BGPS/distance-omnibus}}  Like the theoretical framework for the DPDFs, the \omni\ code base is extremely flexible, and additional prior DPDF methods may be developed and implemented through the use of text configuration files.  The user has control over which DPDFs are computed as well as Galactic properties such as rotation curve and $R_0$.  The code is designed to run nearly autonomously and create the various intermediate data products on-the-fly given the disk locations of the required ancillary data sets.  Presently available in the v1.0 release are the methods presented in EB13 and this paper.


\section{DPDFs AND GALACTIC KINEMATICS}\label{ch3:dpdf}

\subsection{Overview of the DPDF Formalism}

The DPDF formalism introduced in EB13 provides a powerful yet flexible technique for applying a diverse set of ancillary data to the problem of estimating distances to continuum-detected molecular cloud structures in the Galactic plane.  In this Bayesian framework, the posterior DPDF for an object or position on the sky is given by 
\beqn\label{eqn:ch3_dpdf}
\mathrm{DPDF}(d_{_\sun}) = \mathcal{L}(v_\mathrm{LSR},l,b;d_{_\sun})~\prod_i P_i(l,b;d_{_\sun})~,
\eeqn
where $\mathcal{L}(v_\mathrm{LSR},l,b;d_{_\sun})$ is the kinematic distance likelihood function,\footnote{In EB13, this was referred to as DPDF$_\mathrm{kin}$.  We shift to the present notation to distinguish between the kinematic distance likelihood function and the prior DPDFs used to resolve the KDA.} and the $P_i(l,b;d_{_\sun})$ are prior DPDFs based on ancillary data, principally used to resolve the KDA for objects in the inner Galaxy.  The posterior DPDF is normalized to unit integral probability.  The kinematic distance likelihood function is obtained from the marginalization over \vlsr\ of the product of the projected rotation curve along ($\ell,b$) as a $f($\dsun,\vlsr) with the molecular line emission spectrum (itself a function of \vlsr), and is generally double-peaked for Galactocentric positions within the solar circle (\ie $R_\mathrm{gal} < R_0$).  Any number of priors may be applied to resolve the KDA for a given molecular cloud structure, and any given prior may be restricted in its applicability (limited longitude coverage, etc.).

While the true power of the DPDF lies in being a continuous function of \dsun, there are instances where a single-value distance estimate is desired.  Two primary distance estimates may be derived from a DPDF: the maximum-likelihood distance (\dml), and the weighted-average first-moment distance (\dbar).  The \dml\ is simply the distance corresponding to the largest probability in the posterior DPDF, while \dbar\ is computed via $\overline{d} = \int_0^\infty d_{_\sun}\ \mathrm{DPDF}\ \mathrm{d}(d_{_\sun})$.  If the DPDF is well-constrained to have a single peak, \dml\ and \dbar\ will be nearly equivalent, but in cases where the KDA resolution is not well-constrained, these distance estimates may be substantially different and \dbar\ becomes a biased estimator of the distance while \dml\ becomes less reliable.  For well-constrained distances away from the tangent point we recommend \dml, but \dbar\ represents a better distance estimate for objects near the tangent point due to the complex, multimodal structure of $\mathcal{L}(v_\mathrm{LSR},l,b;d_\sun)$ in these regions.

\subsection{Galactic Kinematics}\label{dpdf:kin}
\subsubsection{The Galactic Rotation Curve}\label{res:rotcurve}

\begin{deluxetable*}{lcccc}
  \tablecolumns{5}
  \tablewidth{0pc}
  \tabletypesize{\footnotesize}
  \tablecaption{Galactic Rotation Curve Parameters\label{table:mw}}
  \tablehead{
    \colhead{Parameter} & \colhead{\citeauthor{Reid:2009}} & \colhead{Used in} & \colhead{\citet{Reid:2014}} & \colhead{Used} \\
    \colhead{} & \colhead{\citeyearpar{Reid:2009}} & \colhead{EB13\tablenotemark{a}} & \colhead{Model A5} & \colhead{Here}
  }
  \startdata
  $R_0$ (kpc)                            & $8.4\pm0.6$ & $8.51$ & $8.34 \pm 0.16$ & $8.34$ \\ 
  $\Theta_0$ (\kms)                      & $254\pm16$ & $244$  &  $240 \pm 8$    & $240$ \\
  $\frac{d\Theta}{dR}$ (\kms\ kpc$^{-1}$) & $0$ & $0$   &  $-0.2 \pm 0.4$  &  $0$ \\
  \\
  $U_\sun$ (\kms) & $10.0 $ & $11.1$   &  $10.7 \pm 1.8$  &  $10.7$ \\
  $V_\sun$ (\kms) & $5.2$ & $12.24$  &  $15.6 \pm 6.8$  &  $15.6$ \\
  $W_\sun$ (\kms) & $7.2$ & $7.25$   &  $8.9 \pm 0.9$   &  $8.9$ \\
  \\
  $\overline{U_s}$ (\kms) & $0$ &  $0$   &  $2.9 \pm 2.1$   &  $2.9$ \\
  $\overline{V_s}$ (\kms) & $-15$ &  $-6$  &  $-1.6 \pm 6.8$  &  $0$
  \enddata
  \tablenotetext{a}{Values from M. Reid (2011, private communication), \citet{Schonrich:2010}}
\end{deluxetable*}

\begin{figure}[!t]
        \centering
        \includegraphics[width=3.1in]{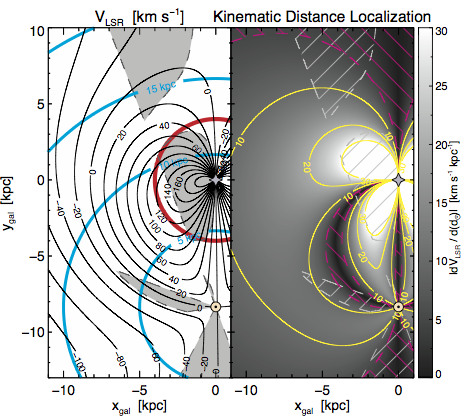}
        \caption[Galactic kinematics for distance determination.]{Galactic kinematics for distance determination.  \emph{Left}:  Contours mark lines of constant projected \vlsr\ based on the (flat) rotation curve of model A5 from \citet{Reid:2014}.  Regions for which kinematic distances are not computed (kinematic avoidance zones) are shown in gray with dashed borders.  The Sun (sun symbol) and Galactic center (star) are marked for reference.  Cyan contours mark \dsun\ in 5-kpc intervals, and the red circle identifies $R_\mathrm{gal} = 4$~kpc, inside which the Galactic bar dominates the kinematics.  \emph{Right}: The magnitude of the derivative of \vlsr\ with respect to \dsun\ across the Galactic plane (grayscale).  Magenta hashing identifies regions where $|dv_{_\mathrm{LSR}} / d(d_{_\sun})| \leq 5$~\kms~kpc$^{-1}$ (\ie a projected peculiar motion of 5~\kms\ will yield a $\geq 1$~kpc distance error).  Kinematic avoidance zones are marked by gray hashing, and yellow contours mark constant values of $|dv_{_\mathrm{LSR}} / d(d_{_\sun})|$.}
        \label{fig:usbar}
\end{figure}

The choice of Galactic rotation curve can have a significant effect on derived kinematic distances.  Over the years, various approaches have been used for deriving the rotation curve of the Milky Way, a decidedly complex endeavor given our location in the Galactic plane.  \citet{Clemens:1985} computed a rotation curve using measured tangent velocities as a function of Galactic longitude from the Massachusetts-Stony Brook Galactic plane CO survey \citep{Sanders:1985}.  Similarly, \citet{Brand:1993} used tangent-point data from 21-cm \HI\ surveys and \HII\ region distance and velocity measurements to construct a rotation curve.  Both curves have been widely used to compute kinematic distances for molecular cloud clumps across the Galactic plane.

In addition to determining the spatial structure of the Milky Way, the BeSSeL survey has used their 6-dimensional trigonometric parallax data (position, velocity) to fit Galactic rotation curve models \citep[\cf][]{Reid:2009}.  Unlike curves based on tangent-point data that require assumed values for $R_0$ and $\Theta_0$ (the radius of the solar circle, and rotational speed of that circular orbit, respectively), the fits to parallax data aim to independently measure these fundamental Galactic parameters.  \citet{Reid:2014} fit a relatively simple rotation curve with speed $\Theta(R) = \Theta_0 + \frac{d\Theta}{dR}(R-R_0)$ to the maser parallax data, and applied different sets of priors on certain model parameters to reflect varying confidence in the prior information.  The fit parameters allowed to vary are the three circular rotation curve parameters ($R_0,\Theta_0,\frac{d\Theta}{dR}$), the values of the solar peculiar motion ($U_\sun, V_\sun, W_\sun$), and the mean deviant (non-circular) motions of high-mass star forming regions (HMSFRs) towards the Galactic center ($\overline{U_s}$) and in the direction of Galactic rotation ($\overline{V_s}$).  As discussed by those authors, $\overline{U_s}$ and $\overline{V_s}$ may be thought of as a first approximation of the kinematic effects (\ie non-circular motion) of spiral structure.

We adopt model A5 of \citet{Reid:2014} for the computation of kinematic distances in the DPDF formalism.  This model was fit using a subset of the parallax measurements where outliers have been excluded, and includes loose priors for the $ V_\sun$ component of the solar motion and the mean deviant motions of HMSFRs, reflecting the large uncertainties surrounding measurements of these values.\footnote{There has been some controversy in the literature about the values of the solar peculiar motion and the counterrotation ($\overline{V_s} < 0$) of HMSFRs \citep[\cf][]{Reid:2009,McMillan:2010,Schonrich:2010,Honma:2012}.  While both $\Theta_o + V_\sun$ and $V_\sun - \overline{V_s}$ are highly constrained, the individual parameters are highly correlated \citep{Reid:2014}.}  The resulting fit parameters are listed in Table~\ref{table:mw}, along with the corresponding values used here and in previous works.

For simplicity, we apply a strictly flat rotation curve, consistent with the measured $\frac{d\Theta}{dR} = -0.2\pm0.4$~\kms~kpc$^{-1}$, and we do not apply a correction for the counterrotation of HMSFRs ($\overline{V_s}$), also consistent with the values from model A5.  The nonzero mean source motion towards the Galactic center ($\overline{U_s}$) from model A5 is included (see the derivation of the kinematic distance likelihood function in Appendix~\ref{app:kdist}), and results in a distance adjustment of $\lesssim 0.2$~kpc for the vast majority of BGPS V2 sources.  Contours in Figure~\ref{fig:usbar} (\emph{left}) illustrate the projected \vlsr\ derived from model A5 as a function of Galactocentric position for the northern Galactic plane.
\subsubsection{Kinematic Avoidance Zones}\label{res:kaz}

Use of kinematic distances for molecular cloud structures is predicated on two principal assumptions: (1) the gas being observed is moving in a circular orbit around the Galactic center, and (2) that small deviant motions of the gas about the true \vlsr\ all map to a narrow set of heliocentric distances.  Regions of the Galaxy that violate these assumptions must be excluded from kinematic distance calculations, as derived distances may be off by a factor of two or more \citep[\cf][]{Reid:2009}.  We label these regions ``kinematic avoidance zones.''

Toward the center of the Milky Way, the long Galactic bar at $R_\mathrm{gal} \lesssim 3-4$~kpc \citep[][]{Fux:1999,RodriguezFernandez:2008,Reid:2014} introduces strong radial streaming motions of the gas, and care must be exercised to utilize kinematic information only for sources outside the influence of these orbits.  This amounts to excluding much of $|\ell| \lesssim 20\degr$.  To quantify this exclusion, EB13 (their Figure~5) defined two regions in the longitude-velocity diagram for which $\mathcal{L}(v_\mathrm{LSR},l,b;d_\sun)$ is not computed.  For $7\fdg5 \leq \ell \leq 21\degr$,\footnote{The lower longitude limit on this region is the lowest $\ell$ for which BGPS spectroscopic observations were made.  Significant blending of structures in the $\ell-v$ diagram at lower longitude make use of any kinematic information challenging at best.} the upper exclusion region is bounded by \vlsr~= (3.33~\kms)$\times\ell(\degr) + 15$~\kms, and includes the higher-velocity gas inside the bar.  The lower region excludes the 3-kpc expanding arm, and is bounded by \vlsr~= (2.22~\kms)$\times\ell(\degr) - 16.7$~\kms.  The effective regions in Galactocentric coordinates for these zones, computed for the rotation curve described above, are shown in Figure~\ref{fig:usbar} (\emph{left}) as gray areas towards the Galactic center.  The allowed velocities in this longitude region correspond to the Scutum-Centarus arm feature \citep[\cf][]{Dame:2011}.

Toward the Galactic anti-center ($\ell = 180\degr$), orbital circular motion is nearly perpendicular to the line of sight, making the projection of \vlsr\ onto the rotation curve highly subject to small peculiar motions of gas.  We therefore define the region $\ell = 180\degr \pm 20\degr$ as an additional kinematic avoidance zone (see Figure~\ref{fig:usbar}), and do not compute kinematic distance likelihoods for objects in this area.

Finally, the portions of the Galaxy nearly parallel to the Sun's motion (\ie $\ell = 90\degr$ and $270\degr$) present fairly flat \vlsr(\dsun) over the span of some 5~kpc.  Peculiar motions are very likely to produce wildly inaccurate kinematic distances, as is the case for object G075.76+00.33 (see \S \ref{prior:maser}).  The flatness of the projected rotation curve (\ie $dv_{_\mathrm{LSR}} / d(d_{_\sun}) \leq 5$~\kms~kpc$^{-1}$; magenta hashing in Figure~\ref{fig:usbar} (\emph{right})) is not symmetric about $\ell = 90\degr$ and $270\degr$, but instead follows the tangent circle toward the Galactic center.  The expected virial motion within a HMSFR is $\approx 5$~\kms, and distance errors of $\gtrsim 1$~kpc due to virial motions are not desirable.  At $\ell \lesssim 70\degr$, this region around the tangent point becomes narrower and therefore acceptable.  The rotation curve derivative is steep enough for $\ell \gtrsim 100\degr$ that the kinematic avoidance zone should only be defined over $70\degr \leq \ell \leq 100\degr$.  In this range, however, the small $dv_{_\mathrm{LSR}} / d(d_{_\sun})$ occurs only at small heliocentric distance (and consequently \vlsr~$\sim 0$~\kms).  Examination of the $\ell-v$ diagram reveals that the Cygnus X region ($\ell \approx 80\degr$) extends from about $-15$~\kms~$\leq$ \vlsr~$\leq 20$~\kms\ (falling in the avoidance zone), with the better-defined Perseus and Outer arms visible at more negative \vlsr.  To fully encompass the Cygnus X region as a kinematic avoidance zone, we therefore limit the kinematic avoidance zone here to \vlsr~$\geq -15$~\kms, and the resulting region in Galactocentric coordinates is visible in Figure~\ref{fig:usbar}.  The corresponding zone near $\ell = 270\degr$ would need to be defined based on gas kinematics near the Carina tangent.


\section{MOLECULAR CLOUD CLUMP \thco\ VELOCITY EXTRACTION}\label{ch3:method}

\subsection{Comparison of GRS \thco\ Spectra with Dense Gas Tracers}\label{meth:comp}

The most straightforward method for using CO data to assign \vlsr\ to molecular cloud clumps is to choose the brightest emission peak along the line of sight.  This has been used by various groups \citep[\cf][]{Russeil:2011,Eden:2012} in lieu of observations of molecular line transitions with higher \neff\ that trace the denser gas associated with molecular cloud clumps.  While this process generally selects the dense material seen in dust continuum emission, areas of warmer diffuse gas or multiple molecular cloud clumps along a line of sight may lead to an incorrect \vlsr\ assignment.

\begin{figure}[!t]
  \centering
  \includegraphics[width=3.1in]{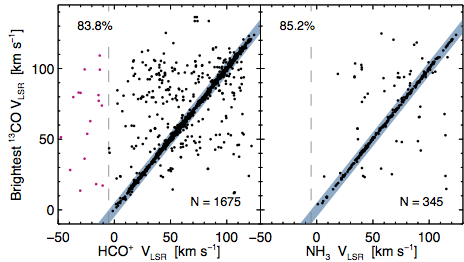}
  \caption[Comparison of the \vlsr\ of the brightest \thco\ emission feature from the GRS data towards a BGPS object compared to the \hcop(3-2) velocities of \citet{Shirley:2013} and \nhhh(1,1) velocities of \citet{Dunham:2011c}.]{Comparison of the \vlsr\ of the brightest \thco\ emission feature from the GRS data towards a BGPS object compared to the (\emph{left}) \hcop(3-2) velocities of \citet{Shirley:2013} and (\emph{right}) \nhhh(1,1) velocities of \citet{Dunham:2011c}.  The vertical gray dashed line in each panel represents the lower \vlsr\ bound of GRS observations.  Objects depicted in pink (\vlsr~$< 5$~\kms) would, therefore, not be detectable in the GRS data.  The light blue region marks the $|\Delta$\vlsr$| \leq 5$~\kms\ used to compute the matching statistic shown at the top of each panel.}
  \label{fig:bright_co}
\end{figure}

To illustrate the challenges of simply choosing the brightest \thco\ peak for BGPS objects, Figure~\ref{fig:bright_co} compares that velocity with available \hcop(3-2) (S13) and \nhhh(1,1) \citep{Dunham:2011c} observations.  The extracted \vlsr\ from \thco\ agrees with the dense-gas value to within 5~\kms\ $\sim85\%$ of the time.  For the purposes of this comparison, the GRS data cubes were averaged over the angular extent of the BGPS source to generate a composite \thco\ spectrum; S13 show that using the single GRS spectrum at the location of peak $\lambda = 1.1$~mm flux density yields an agreement rate $\sim 81\%$.

The GRS was designed to study the chemistry and kinematics of the large peak in molecular gas density about halfway between the Sun and the Galactic center, the so-called ``5-kpc ring'' \citep{Clemens:1988}.  Because there is comparatively little CO emission on the far side of the Galaxy beyond the solar circle behind this feature in the first quadrant \citep[\cf][]{Clemens:1988,Dame:2001}, the GRS did not observe negative \vlsr.  Continuum surveys such as the BGPS detect the optically thin dust within molecular cloud clumps, and so are sensitive to objects on the far side of the Galaxy beyond the solar circle, embedded in what CO is present at those locations (see \S\ref{res:lv}).  Several such objects are illustrated in Figure~\ref{fig:bright_co} as pink dots with \hcop(3-2) \vlsr~$< -5$~\kms.  Of the 1,846 BGPS objects with detected dense-gas \vlsr\ in the GRS overlap region, however, only 18 are beyond the solar circle.

\subsection{Morphological Spectrum Extraction Technique}\label{meth:onoff}

\subsubsection{General Description}

\begin{figure*}[!t]
  \centering
  \includegraphics[width=3.1in]{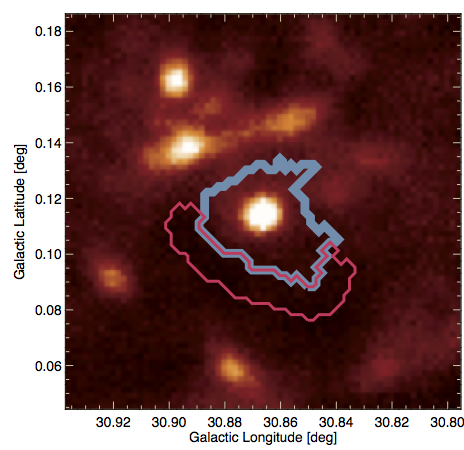}
  \includegraphics[width=3.1in]{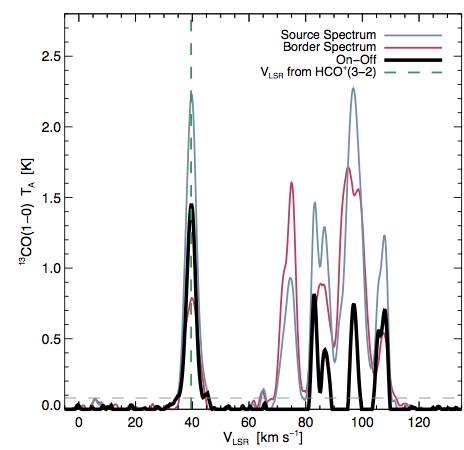}
  \caption[Demonstration of the morphological spectrum extraction technique for example source G030.866+00.115.]{Demonstration of the morphological spectrum extraction technique for example source G030.866+00.115.  \emph{Left}: The image depicts the BGPS V2 flux density on a linear scale from $-0.10$ to $1.40$~Jy beam$^{-1}$.  The blue contour is the Bolocat source outline, as described in \citet{Rosolowsky:2010}, and the pink contour depicts the off-source region, one GRS resolution element in width, excluding pixels associated with the continuum source to the lower right.  \emph{Right}: The filtered \thco(1-0) spectra for this object.  Blue and pink spectra represent the $T_\mathrm{on}(v)$ and $T_\mathrm{off}(v)$, respectively, while black represents \tsource, used to estimate the \vlsr\ of the denser gas associated with the molecular cloud clump.  The vertical green dashed line marks the \hcop(3-2) \vlsr\ from S13.}
  \label{fig:grs_onoff}
\end{figure*}

While the $\sim85\%$ velocity matching rate for the brightest \thco\ feature along a line of sight is encouraging, the low \neff\ of \thco(1-0) means that observed emission is not uniquely tied to molecular cloud clumps.  A large patch of warm, diffuse gas or changes in the excitation temperature may produce an emission feature that outshines all others along a given line-of-sight.  To mitigate this effect, we developed a technique for using millimeter continuum data as prior information for extracting a velocity spectrum for the molecular cloud structure of interest.  The 46\arcsec\ resolution of the GRS is similar enough to that of the BGPS that we may assume both surveys are sensitive to similar structures.

To leverage the continuum data's prior information, we use a morphological spectrum extraction technique to isolate the contribution to the \thco\ emission from a particular molecular cloud structure from its more diffuse envelope \citep{Rosolowsky:2010a}.  The extracted spectrum is computed as $T_\mathrm{source}(v) = T_\mathrm{on}(v) - T_\mathrm{off}(v),$ where $T_\mathrm{on}(v)$ is the average of the GRS spectra at the location of BGPS pixels within the Bolocat label masks \citep[see][]{Rosolowsky:2010}, weighted by BGPS flux density.  Each BGPS pixel is assigned the \thco\ spectrum of the nearest ($\ell,b$) GRS pixel; the same \thco\ spectrum (22\arcsec\ pixels) may therefore be assigned to more than one BGPS pixel (7\farcs2).  The off-source spectrum is computed as the unweighted average of the GRS spectra assigned to BGPS pixels within one GRS resolution element of the Bolocat outline, \emph{excluding} pixels assigned to another Bolocat object.  This exclusion avoids subtracting relevant emission associated with the dense gas of neighboring catalog objects, leaving $T_\mathrm{off}(v)$ contributions only from lines of sight penetrating the more diffuse CO envelope.  The 46\arcsec\ width of the off region corresponds to $\sim 1.1$~pc at \dsun~$= 5$~kpc, well within the size of GRS cataloged clouds \citep{RomanDuval:2010}.  A small handful of BGPS objects ($N \sim 10$) are ``landlocked'' (\ie completely surrounded by other Bolocat sources), and no $T_\mathrm{off}(v)$ may be computed.  For these sources, $T_\mathrm{on}(v)$ is used in place of \tsource, and a flag is set.

We emphasize that the angular transfer functions of ground-based (sub-)millimeter observations make this type of extraction possible, as atmospheric subtraction algorithms necessarily remove large-scale diffuse dust emission, leaving well-defined sources to delineate on- and off-source regions.  Application of this technique to space-based data (\eg Hi-GAL) would likely require filtering of the data to remove Galactic cirrus emission, as the cirrus on scales larger than the BGPS sensitivity can contribute a factor of two to the derived dust column density for clump-scale objects \citep[][]{Battersby:2011}.  To verify the correlation between BGPS-detected sources and physically meaningful structures, we compared the maximum angular scales recoverable by the BGPS against spectral-line mapping surveys, which recover emission at all spatial scales larger than the beam size.  The \water\ Southern Galactic Plane Survey \citep[HOPS;][]{Walsh:2011} observed several molecular transition lines around 24~GHz, mapping nearly 700 molecular cloud clumps in \nhhh(1,1) \citep[][]{Purcell:2012}.  Comparing the maximum solid angle the BGPS recovers ($19.6\arcmin^2$; $\theta_\mathrm{max} \approx 300\arcsec$; G13) with the cloud solid angles from \citeauthor{Purcell:2012} shows that nearly 80\% of the HOPS clumps would be fully recovered by the BGPS, with the remaining being somewhat truncated by the angular transfer function.  Therefore, ground-based dust-continuum data selects nearly all of the region visible in \nhhh(1,1) emission brighter than $\approx 0.2$~K \citep[the average root-mean-squared noise temperature of the HOPS \nhhh(1,1) spectra;][]{Purcell:2012}, and is appropriate for using as a prior for extracting \thco\ spectra for molecular cloud clumps. 

The averaged \thco\ spectra were filtered using a kernel constructed to match the typical widths of \thco\ features (FWHM~$\approx 2-5$~\kms) to conserve the relative widths and heights of spectral line features while significantly reducing the noise.  This allows for the detection of spectral features in \tsource\ with peak intensity less than the nominal GRS noise level.  The resulting spectrum is clipped to contain only positive (emission) values.  Figure~\ref{fig:grs_onoff} illustrates the morphological spectrum extraction process for object G030.866+00.115.  In the left panel, the blue contour delineates the source, with the weighted-average GRS spectrum shown in blue in the right panel.  The ``Off'' region (pink outline) excludes BGPS pixels associated with the sources above and to the right, and the averaged GRS spectrum from this region is shown in pink in the right panel.  While there is a general correspondence between the peaks of $T_\mathrm{on}(v)$ (blue) and $T_\mathrm{off}(v)$ (pink), the complicated nature of the emission in this region makes identifying a single or dominant peak from $T_\mathrm{on}(v)$ challenging at best.

\begin{figure}[!t]
  \centering
  \includegraphics[width=3.1in]{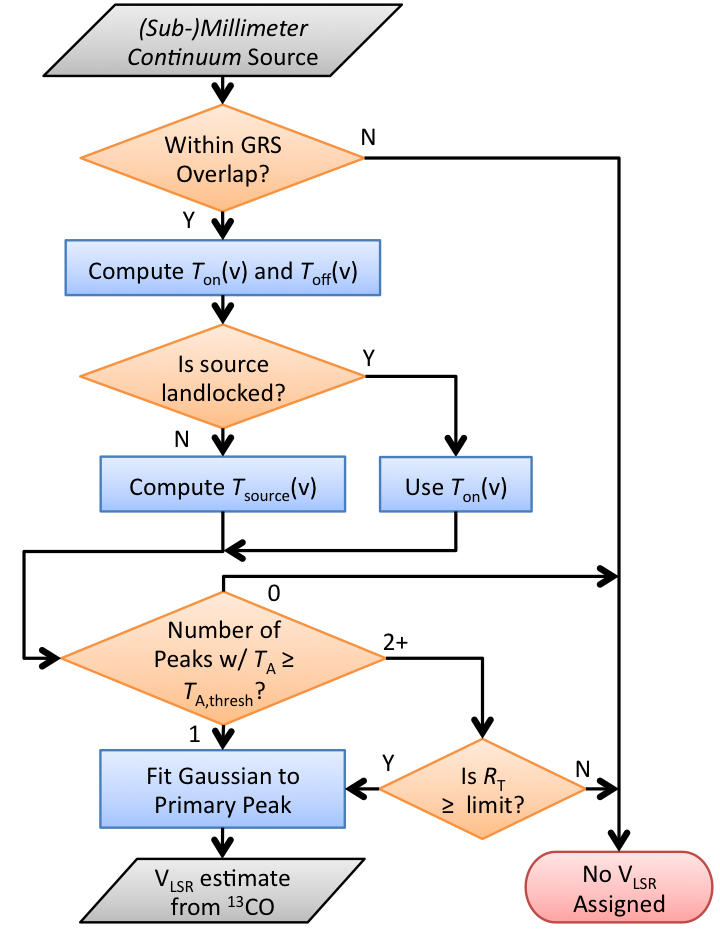}
  \caption[Flow chart for the assignment of \vlsr\ from the morphologically extracted \thco\ spectrum for a given molecular cloud structure.]{Flow chart for the assignment of \vlsr\ from the morphologically extracted \thco\ spectrum for a given molecular cloud structure.}
  \label{fig:grs_flow}
\end{figure}

\subsubsection{Extracting \vlsr\ from \tsource}\label{meth:tsource}

This new technique effectively yields a pointed catalog of ``position-switched'' \thco(1-0) spectra, where the ``reference position'' is carefully chosen to include the diffuse envelope surrounding the molecular cloud clump.  Like any such catalog, detection thresholds and flagging are required to produce a reliable set of \vlsr\ for kinematic distance computation.  Two parameters control the quality and quantity of extracted spectra: the minimum $T_A$ threshold for peaks in \tsource\ ($T_{A,\mathrm{thresh}}$), and the degree to which a single \vlsr\ peak dominates the final spectrum.  We parameterize this latter quantity as the minimum ratio of the $T_A$ of the primary peak to that of the secondary peak (when present), or $R_T = T_{A(1)} / T_{A(2)}$.  For the example source in Figure~\ref{fig:grs_onoff}, the primary peak in \tsource\ is near \vlsr~= 40~\kms, and the secondary peak near \vlsr~= 83~\kms, with $R_T \approx 1.8$.

\begin{figure}[!t]
  \centering

  \includegraphics[width=3.1in]{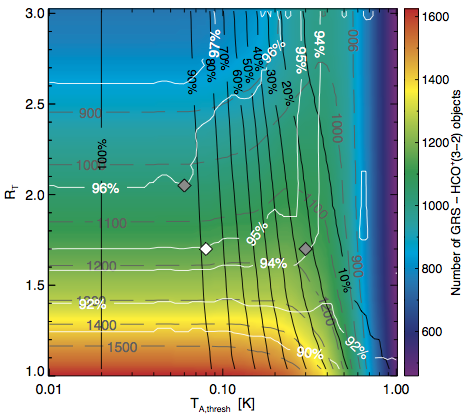}
  \caption[Optimization of the $T_{A,\mathrm{thresh}}$ for detection and $R_T$ limit for multiply peaked spectra.]{Optimization of the $T_{A,\mathrm{thresh}}$ for detection and $R_T$ limit for multiply peaked spectra.  Background color scale and gray dashed contours denote the number of BGPS V2 objects that possess both \hcop(3-2) and morphologically extracted \thco\ velocities ($N_\mathrm{comp}$).  White contours mark the fraction of $N_\mathrm{comp}$ whose \vlsr\ from the two sources agree to better than 5~\kms.  Black contours show the fraction of $N_\mathrm{comp}$ whose morphologically extracted spectrum contains more than one peak above $T_{A,\mathrm{thresh}}$.  The white diamond indicates the chosen values of $T_{A,\mathrm{thresh}}$ and $R_T$ while the gray diamonds are discarded choices (see text).}
  \label{fig:grs_opt}
\end{figure}

The algorithm for \vlsr\ extraction from \tsource\ is illustrated in the flow chart of Figure~\ref{fig:grs_flow}.  For objects within the GRS coverage region, \tsource\ is examined for any peak above $T_{A,\mathrm{thresh}}$.  Next, the number of contiguous regions above $T_{A,\mathrm{thresh}}$ is counted; for two or more independent peaks, $R_T$ is computed.  There are three points in the process where a null \vlsr\ may be returned: (1) the molecular cloud structure is outside the limits of the GRS, (2) there are no peaks in \tsource\ above $T_{A,\mathrm{thresh}}$, and (3) for multiply-peaked spectra there is no clearly dominant peak (\ie $R_T <$ limit).  For the remaining multiply-peaked and all single-peaked spectra, a Gaussian is fit to the primary peak.

To determine the optimal values of these parameters for \vlsr\ extraction, we computed the following as functions of the two tunable parameters: (1) the number of objects ($N_\mathrm{comp}$) having both a valid \thco\ \vlsr\ and \hcop\ \vlsr\ from S13, (2) the fraction of $N_\mathrm{comp}$ whose \thco\ and \hcop\ velocities agree to $\leq 5$~\kms\ (the approximate virial motion within HMSFRs), and (3) the fraction of $N_\mathrm{comp}$ whose \thco\ spectra contain multiple peaks above $T_{A,\mathrm{thresh}}$.  Based on the topography of the parameter space, as shown in Figure~\ref{fig:grs_opt}, we chose values of $T_{A,\mathrm{thresh}} = 0.08$~K, and $R_T = 1.70$ (white diamond) to best balance the number of BGPS V2 objects in the GRS - \hcop\ overlap with the fraction of those sources whose \vlsr\ agree to better than 5~\kms.  

We note that at the selected values for $T_{A,\mathrm{thresh}}$ and $R_T$, nearly 90\% of the morphologically extracted spectra are multiply-peaked (black contours), yet the velocity-agreement fraction is virtually unchanged down to $\sim 30\%$ multiplicity.  Two discarded values are shown as gray diamonds for comparison.  The diamond at ($T_{A,\mathrm{thresh}}, R_T$) = (0.30~K, 1.7) has a threshold at approximately twice the rms noise of the GRS data cubes, and follows the same $N_\mathrm{comp} \approx 1150$ contour the white diamond is on.  The matching rate at this point, however, is 94\%.  A point with a higher matching percentage (96\%) is the gray diamond at ($T_{A,\mathrm{thresh}}, R_T$) = (0.06~K, 2.05).  While this is one percent better than the white diamond, there are $\approx 100$ fewer objects for which a morphological \vlsr\ may be extracted from the \thco\ data.  The white diamond was ultimately a choice made to balance $N_\mathrm{comp}$ with the matching percentage.

\subsubsection{Comparison with \hcop(3-2)}\label{meth:comp2}

\begin{figure}[!t]
  \centering
  \includegraphics[width=3.1in]{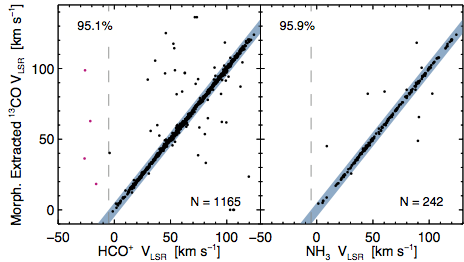}
  \caption[Morphologically extracted \thco\ spectrum \vlsr\ comparison with dense-gas tracers.]{Morphologically extracted \thco\ spectrum \vlsr\ comparison with dense-gas tracers.  Colors and lines as in Figure~\ref{fig:bright_co}.  \emph{Left}: Velocity comparison for the $N=1,165$ BGPS V2 objects with a valid morphologically extracted \thco\ spectrum and \hcop(3-2) \vlsr\ from \citet{Shirley:2013}.  \emph{Right:} Velocity comparison for the $N=242$ BGPS V2 objects with a valid morphologically extracted \thco\ spectrum and \nhhh(1,1) \vlsr\ from \citet{Dunham:2011c}.}
  \label{fig:13co_hcop}
\end{figure}

With the optimized parameter values for $T_{A,\mathrm{thresh}}$ and $R_T$, there are 1,165 objects which possess both an \hcop(3-2) detection and a valid morphologically extracted spectrum from the GRS \thco\ data.  The velocity comparison is shown in Figure~\ref{fig:13co_hcop}, which echoes Figure~\ref{fig:bright_co}, except that the correspondence has grown to $\approx 95\%$.  As with Figure~\ref{fig:bright_co}, pink circles at the left of the plot mark continuum-detected molecular cloud structures beyond the solar circle ($R_\mathrm{gal} > R_0$) that also have detectable \thco\ emission in the foreground.  It is likely these objects would have the correct \vlsr\ assigned (to 95\% confidence) if the GRS had extended velocity coverage to negative \vlsr.  The histogram of the velocity difference for all points in Figure~\ref{fig:13co_hcop} is well-fit by a Gaussian with FWHM~$\approx 1.1$~\kms, the channel width of the \hcop\ spectra (S13).

\subsection{The \thco\ Kinematic Distance Likelihood Function}\label{meth:dpdf}

Using the optimized parameters for computing a morphologically extracted spectrum from the GRS \thco\ data, a kinematic distance likelihood function may be computed for BGPS sources with a valid \thco\ spectrum.  To maximize the use of kinematic information generated from this technique, we use the \tsource\ directly to compute $\mathcal{L}(v_\mathrm{LSR},l,b;d_\sun)$, following the process described in EB13.  Examination of Figure~\ref{fig:grs_onoff}, however, reveals many small features in the extracted spectrum below the $T_A$ threshold that are artifacts of the technique in addition to the stronger secondary peaks at 80~\kms~$\leq$ \vlsr~$\leq 110$~\kms.  To eliminate these extraneous bumps, which would appear as spurious small peaks in the DPDF, we mask \tsource\ to include only the region \vlsr~$\pm~3\sigma$ from the Gaussian fit to the primary peak before multiplying it by the rotation curve probability density function.  For the example source in Figure~\ref{fig:grs_onoff}, this preserves the main peak, while eliminating probability associated with the secondary peaks.


\section{CATALOG-BASED PRIOR DPDFs}\label{ch3:priors}


As large-scale surveys of Galactic star formation have been published over the last several years, they are often accompanied by catalogs of distance estimates.  While some methods are more robust and/or accurate than others, these catalogs offer additional anchor points for use with the DPDF formalism.  With the relatively small present suite of prior DPDFs, the use of literature catalogs can expand the types of distance methods available for use with detected molecular cloud structures.  In this section, we describe two literature catalogs of objects associated with sites of massive star formation (trigonometric parallax measurements of masers in \S\ref{prior:maser}, and robust KDA resolutions for \HII\ regions in \S\ref{prior:hrds}), and define a means for associating distances from those catalogs with BGPS molecular cloud structures (\S\ref{prior:define}).

\subsection{Trigonometric Parallax Measurements}\label{prior:maser}

\begin{figure*}[!t]
  \centering
  \includegraphics[width=3.9in]{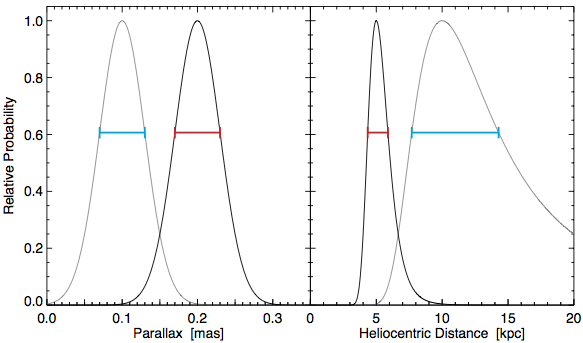}
  \includegraphics[width=2.3in]{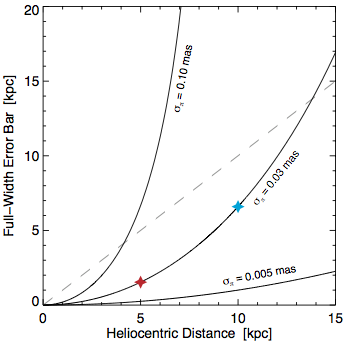}
  \caption[Example of DPDF$_\mathrm{px}$ construction from trigonometric parallax measurements.]{Example of DPDF$_\mathrm{px}$ construction from trigonometric parallax measurements.  \emph{Left}: Gaussians in parallax with $\mu_\mathrm{px} = 0.20$~mas ($\mu_\mathrm{px} = 0.10$~mas) and $\sigma_\mathrm{px} = 0.03$~mas are plotted in black (gray).  The full-width error bars (\ie $\mu\pm\sigma$) are shown in red (cyan).  \emph{Middle}: Interpolation of the parallax functions onto a linear distance scale, where \dsun~$= 1/\pi_s$.  The corresponding full-width error bars are shown in red (cyan).  \emph{Right}: Size of the full-width distance error bar as a function of parallax measurement uncertainty and heliocentric distance.  The curves shown represent the lower bound on the uncertainty ($\sigma_\pi = 0.005$~mas), a typical value ($\sigma_\pi = 0.03$~mas), and a poorly-constrained parallax measurement ($\sigma_\pi = 0.10$~mas).  The red and cyan stars identify the examples at left, and the gray dashed line marks the the 1:1 relationship.}
  \label{fig:px_example}
\end{figure*}

The BeSSeL survey has been conducting VLBI observations of \methanol\ and \water\ masers associated with HMSFRs for the past several years \citep[\cf][and references therein]{Brunthaler:2011,Reid:2014}.  The geometric distances returned by trigonometric parallaxes depend upon neither the choice of Galactic rotation curve nor other assumptions about the structure of the Galactic plane, offering an absolute distance benchmark.  Parallax distances are especially useful within the kinematic avoidance zones described in \S\ref{res:kaz}.  For instance, the $\approx~16$~\kms\ counterrotation of HMSFR G075.76+00.33 yields a kinematic distance of $\approx~5.7$~kpc, apparently in the Perseus arm, but its parallax distance of $3.5\pm0.3$~kpc places it in the Local arm \citep{Xu:2013}.

\begin{deluxetable*}{lccccccc}
  \tablecolumns{8}
  \tablewidth{0pc}
  \tabletypesize{\footnotesize}
  \tablecaption{\HII\ Region KDA Resolutions from the HRDS\label{table:hrds}}
  \tablehead{
    \colhead{Object}  & \colhead{$\ell$} & \colhead{$b$} & \colhead{\vlsr} & \colhead{\dsun} & \colhead{KDA\tablenotemark{a}} & \colhead{QF\tablenotemark{b}} & \colhead{Ref.} \\
    \colhead{Name} & \colhead{(\degr)} & \colhead{(\degr)} & \colhead{(\kms)} & \colhead{(kpc)} & \colhead{Resol.} & \colhead{} & \colhead{}
  }
  \startdata
U23.20+0.00a & $23.200$ & \phs$0.000$ & \phs\phn$22.4$ & 14.00 & F & B & 1 \\
U23.43-0.21 & $23.430$ & $-0.210$ & \phs$101.1$ & \phn6.00 & N & A & 1 \\
U23.96+0.15 & $23.960$ & \phs$0.150$ & \phs\phn$78.9$ & \phn5.00 & N & A & 1 \\
C24.30-0.15a & $24.300$ & $-0.150$ & \phs\phn$55.5$ & 11.70 & F & A & 1 \\
C27.49+0.19 & $27.490$ & \phs$0.190$ & \phs\phn$34.0$ & 12.80 & F & A & 1 \\
G032.272-0.226 & $32.272$ & $-0.226$ & \phs\phn$21.5$ & 12.80 & F & A & 2 \\
C33.42+0.00 & $33.420$ & \phs$0.000$ & \phs\phn$76.5$ & \phn9.40 & F & A & 1 \\
G038.738-0.140 & $38.738$ & $-0.140$ & \phs\phn$60.9$ & \phn9.20 & F & B & 2 \\
G046.948+0.374 & $46.948$ & \phs$0.374$ & \phn$-44.4$ & 16.20 & F & A & 2 \\
G047.094+0.492 & $47.094$ & \phs$0.492$ & \phn$-54.5$ & 17.50 & F & \nodata & 3
  \enddata
  \tablerefs{1: \citet{Anderson:2009a}; 2. \citet{Anderson:2012}, 3. \citet{Bania:2012}}
  \tablenotetext{a}{N = near, F = far, T = tangent}
  \tablenotetext{b}{Quality Factor of the KDA resolution; see \citet{Anderson:2012} for more details.  \citet{Bania:2012} does not assign a QF.}
  \tablecomments{This table is available in its entirety in a machine-readable format in the online journal.  A portion is shown here for guidance regarding its form and content.}
\end{deluxetable*}

To construct the prior DPDF$_\mathrm{px}$ for trigonometric parallax measurements ($\pi_s$) associated with molecular cloud structures, we utilize the latest list of robust parallaxes from BeSSeL and VERA \citep[][Table~1]{Reid:2014}, which contains measurements for 103 sources associated with HMSFRs throughout $-12\degr \leq \ell \leq 241\degr$.  Source ($\ell,b,$\vlsr) from the BGPS V2.1 catalog are compared with the association volumes for each maser in the parallax table, as discussed above.  Because of the ``gold standard'' nature of trigonometric parallax measurements, there is no need to dilute DPDF$_\mathrm{px}$ to allow nonzero probability far from the measured distance.  The prior DPDF is therefore created by constructing a gaussian in parallax space using the value and uncertainty from \citet[][Table~1]{Reid:2014} as the centroid and standard deviation, respectively.  Next, that array is interpolated onto a linear distance array, where \dsun~$= 1/\pi_s$, with points spaced every 0.02~kpc to create a DPDF reflecting the asymmetric nature of parallax distance uncertainties.  Example DPDF$_\mathrm{px}$ for $\pi_s = 0.20 \pm 0.03$ (black) and $\pi_s = 0.10 \pm 0.03$ (gray) are shown in Figure~\ref{fig:px_example}.  The typical parallax uncertainty in the present sample is $\approx 0.03$~mas, which translates to a distance uncertainty of $^{+0.88}_{-0.65}$~kpc at \dsun~= 5~kpc, or a full-width error bar (see \S\ref{res:wc}) of 1.53~kpc.  The right panel of Figure~\ref{fig:px_example} illustrates the dependence of the distance uncertainty on heliocentric distance for three values of parallax uncertainty, showing the rapid increase in the fractional parallax uncertainty as a function of \dsun.

\subsection{\HII\ Region KDA Resolutions}\label{prior:hrds}

Galactic \HII\ regions associated with HMSFRs offer an additional opportunity for generating prior DPDFs for molecular cloud structures.  The \HII\ Region Discovery Surveys \citep[HRDS;][]{Bania:2010,Bania:2012} used the Green Bank Telescope (GBT) and Arecibo Observatory to search for radio recombination lines indicative of ionized gas based on a candidate list compiled principally from mid-infrared and radio continuum data and catalogs.  The resulting collection of previously known and newly discovered \HII\ regions (448 from GBT and 37 from Arecibo) associated with massive star formation provide a sizable catalog from which to assign KDA resolutions for molecular cloud structures \citep[][]{Anderson:2009a,Anderson:2012,Bania:2012}.

KDA resolutions for HRDS sources rely upon the \HI\ self-absorption (HISA) and \HI\ emission~/ absorption (HIE/A) methods \citep[\cf][]{Anderson:2009a,RomanDuval:2009}.  In short, both methods make use of cold \HI\ in molecular clouds absorbing 21-cm emission from a backlighting source.  A stable population of neutral atomic hydrogen is maintained even in the cold, dense regions of molecular cloud clumps through an equilibrium between \htwo\ formation and destruction by cosmic rays \citep{Goldsmith:2007}.  For HISA, cold gas at the near kinematic distance absorbs 21-cm emission from warm gas at the same \vlsr\ at the far kinematic distance.  In the HIE/A  method, 21-cm continuum emission from the \HII\ region is examined for absorption features; absorption at $|v_\mathrm{HII}| < |v_{_\mathrm{absorption}}| < |v_\mathrm{tan}|$ places the \HII\ region at the far kinematic distance, and absorption only at $|v_{_\mathrm{absorption}}| < |v_\mathrm{HII}|$ places it at the near.

The set of 441 \HII\ regions with strong KDA resolution \citep[quality factor A or B; see][]{Anderson:2012} from the various HRDS publications are gathered in Table~\ref{table:hrds} as a reference for computing DPDF$_\mathrm{hrds}$.  Since the HRDS Galactic longitude range is $15\degr \leq \ell \leq 67\degr$, all objects with positive \vlsr\ must be assigned a KDA resolution (objects with \vlsr~$\leq 0$~\kms\ are unambiguously beyond the solar circle).  Because this prior is based on the physically relevant tangent distance (\dtan), we construct DPDF$_\mathrm{hrds}$ as a step function removing probability on the opposite side of \dtan.  An analysis of posterior DPDFs suggests that objects within 1~kpc of \dtan\ should be given a ``tangent'' KDA resolution (\S\ref{res:wc}), so a strict step function at \dtan\ would improperly bias the prior for these objects.  We therefore model DPDF$_\mathrm{hrds}$ with an error function possessing a rolloff width of 1~kpc for objects in Table~\ref{table:hrds} with a ``N'' or ``F'' resolution, and as a Gaussian centered on \dtan\ with $\sigma = 1$~kpc for objects with a ``T'' resolution.

\subsection{Associating Molecular Cloud Structures with Literature Catalog Objects}\label{prior:define}

\subsubsection{Definitions}

Care must be taken when applying distance information from literature catalogs to create a prior DPDF.  The clumpy nature of the interstellar medium and even within GMCs requires defining a volume around literature catalog objects for association with molecular cloud structures.  Given that the three observational dimensions are plane-of-sky and velocity, we define a cylindrical association volume in coordinate-velocity ($\ell,b,$\vlsr) space whose symmetry axis lies along the line of sight to the catalog object.  Proper physical scaling of these cylinders should be based on typical coherent structures in the interstellar medium, namely GMCs.  We use here the collection of GMCs cataloged by the GRS \citep[][]{Rathborne:2009}, which represent the coherent envelopes within which several or many molecular cloud clumps may reside.  The physical properties of these GMCs \citep[as computed by][]{RomanDuval:2010} provide reasonable baselines for fixing the association volume.

The likelihood of a single molecular cloud structure lying within the association volume of more than one catalog object (\eg \HII\ regions) is nonzero.  Distance assignments or KDA resolutions for multiple catalog entries may conflict, so a mechanism for combining information from multiple objects is required.  Because a catalog object lying closest to a molecular cloud structure in coordinate-velocity space is more likely to assign the correct distance, we combine prior information from multiple catalog entires weighted by $1/\xi$, where the non-dimensional distance parameter is computed as
\beqn\label{eqn:xi}
\xi = \left[ \left( \frac{\Delta\theta}{\theta_a}\right)^2 + \left( \frac{\Delta v_{_\mathrm{LSR}}}{\Delta v_a}\right)^4  \right]^{1/2}~.
\eeqn
The quantities $\Delta\theta$ and $\Delta$\vlsr\ are the angular separation on the sky and velocity separation, respectively, between the continuum-identified source and the catalog object, and $\theta_a$ and $\Delta v_a$ are the optimized association volume angular radius and velocity extent, respectively (determined below).  The physical association volume is based on the physical radius of GRS clouds ($R_a$), so $\theta_a$ must be computed individually for each catalog object as $R_a / $\dsun.  The turbulent structure function dictates that $\Delta v \sim R^{1/2}$ \citep{Heyer:2009}, giving rise to the quartic velocity term in Equation~(\ref{eqn:xi}).  Since association is not considered outside the volume defined by $\theta_a$ and $\Delta v_a$, the maximum possible value of $\xi$ is $\sqrt2$.  In the case of competing KDA resolutions for catalog objects with similar $\xi$, the resulting prior DPDF will be relatively unconstrained, accurately reflecting the distance uncertainty given the available information.

\subsubsection{Optimization}\label{res:assoc}

\begin{figure}[!t]
  \centering
  \includegraphics[width=3.1in]{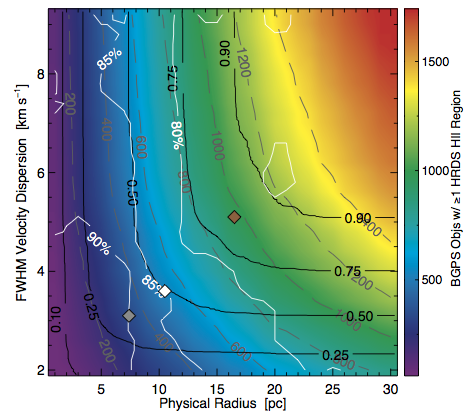}
  \caption[Optimization of the association volume for catalog-based priors, based on HRDS \HII\ regions.]{Optimization of the association volume for catalog-based priors, based on HRDS \HII\ regions.  Background color scale and gray dashed contours denote the number of BGPS V2 objects falling within the association volume of one or more HRDS \HII\ region, given the $R_a$ and $\Delta v_a$ indicated.  Black contours show the bivariate cumulative distribution function for $R$ and $\Delta v$ for 749 GRS clouds from \citet{RomanDuval:2010}, where the values shown indicate the fraction of clouds whose physical properties are \emph{both} smaller than the axes values at that point.  White contours mark the fraction of BGPS sources whose EMAF-derived distances and \HII\ region-derived distances agree to better than 1~kpc.  The white diamond indicates the chosen values of $R_a$ and $\Delta v_a$ while the brown and gray diamonds are discarded choices (see text).}
  \label{fig:grs_prop}
\end{figure}

As with the morphological spectrum extraction technique of \S\ref{ch3:method}, we must balance the number of objects to which this method may apply with some measure of the method's accuracy over the parameter space formed by the physical association volume values ($R_a, \Delta v_a$).  The figure of merit for determining the optimization of this method is the agreement fraction between the heliocentric distance returned for BGPS catalog sources by the EMAF and \htwo\ priors (EB13) versus that of the HRDS prior described above.  While the optimization over association volume is done with respect to GRS clouds (size scale of coherent structure), the EMAFs and \HII\ regions whose distances are compared are most likely substructures of the larger clouds.  Showing that these smaller objects can be associated across GMC-scale distances indicates the validity of this method.

The physical properties of the 749 GRS clouds studied by \citet{RomanDuval:2010} form the parameter space over which this optimization is performed, as they represent coherent molecular structures in ($\ell,b,$\vlsr) space.  The bivariate cumulative distribution function of these quantities for these GRS clouds is shown as black contours in Figure~\ref{fig:grs_prop}, where the value shown at any point represents the fraction of GRS clouds whose ($R_a, \Delta v_a$) are both smaller than that point.  For example, 75\% of GRS clouds have ($R_a, \Delta v_a$) smaller than the line marked 0.75.  The bivariate distribution value is generally smaller for any given point in the plane than the marginalized cumulative distributions for the individual properties.  For instance, the brown diamond marks the 90\sth percentile values for each of $R_a$ and $\Delta v_a$ individually, but the combination lies along the 83\srd percentile contour.

Akin to the color scale of $N_\mathrm{comp}$ in Figure~\ref{fig:grs_opt}, the color scale and gray dashed contours in Figure~\ref{fig:grs_prop} depict the number of BGPS V2 sources associated with one or more HRDS \HII\ region (\S\ref{prior:hrds}).  The number of associated sources grows very rapidly with increasing association volume, and small shifts in ($R_a, \Delta v_a$) can have a factor-of-two effect on the number of sources included.  While large association volumes would assign prior DPDFs to many molecular cloud structures, a regime is quickly reached where that volume far exceeds the physical sizes of GMCs.  Conversely, small association volumes are assured to lie entirely within most GMCs, but the usefulness of the prior becomes limited.

To assess the accuracy of prior DPDFs in resolving the KDA, we compared the heliocentric distance derived using only DPDF$_\mathrm{hrds}$ (with association volume $R_a,\Delta v_a$) as a prior on $\mathcal{L}(v_\mathrm{LSR},l,b;d_\sun)$ using Equation~(\ref{eqn:ch3_dpdf}) with the distance derived using the priors DPDF$_\mathrm{emaf}$ and DPDF$_\mathrm{H_2}$ from EB13 in the same fashion.  This comparison was done over the relevant range of ($R_a, \Delta v_a$) parameter space, given the GRS cloud properties from \citet{RomanDuval:2010}.  At each point in the ($R_a, \Delta v_a$) plane, the comparison was computed using the set of EMAF-idenified BGPS sources associated with one or more \HII\ regions given those values for the association volume.  The rate at which the heliocentric distances from the two methods agree to better than 1~kpc is shown by the white contours in Figure~\ref{fig:grs_prop}.  We used the DPDF$_\mathrm{emaf}$ as the basis for calibrating the literature-catalog association volume prior because it provides a high-accuracy distance estimate compared to GRS cloud KDA resolutions ($\approx 92\%$; EB13) and provides well-constrained distance estimates for more than 700 molecular cloud structures.  With large physical (three-dimensional) separation between BGPS sources and cataloged \HII\ regions, the distance agreement rate falls from $\gtrsim 90\%$ to $\approx 80\%$ while the number of possible sources increases many-fold.  The fact that the distance-agreement rate does not fall below 80\% likely indicates the expected existence of larger structure within the Galactic plane beyond the scale of GMCs.  For comparison, if the objects with near and far KDA resolutions in the V2 distance catalog (see \S\ref{res:v2}) are randomly reassigned and those at the tangent point remain unchanged, the distance agreement would be only  $\approx 64\%$.

As with optimizing the morphological spectrum extraction technique of \S\ref{ch3:method}, choosing optimum values of ($R_a, \Delta v_a$) requires balancing distance resolution accuracy with the number of objects to which the method may apply.  To encompass the physical properties of the bulk of GRS clouds, we focused on the 50\sth percentile contour of the bivariate cumulative distribution function, and settled on $(R_a,\Delta v_a)$ = (10.5~pc, 3.6~\kms) as the best balance between distance agreement rate and number of objects included (white diamond).  With these values, 15\% of the EMAF- and \HII\ region-derived distances disagree with each other.  The contrasting physical conditions in these tracers of star formation (cold, dense, starless gas versus hot bubbles around young stars) accounts for much of the difference, given the assumption of smooth Galactic 8-\micron\ emission in DPDF$_\mathrm{emaf}$.  Furthermore, assuming each prior has an independent distance-assignment success rate of 92\% (EB13) regardless of physics, the comparison of distances should agree at a rate of $(0.92)^2 \approx 0.85$, the rate indicated by the white diamond.

Other diamonds in Figure~\ref{fig:grs_prop} represent sub-optimal values.  The brown diamond (mentioned above) has a relatively large association volume and encompasses nearly 1,000 BGPS objects, but is larger than 83\% of GRS clouds and has a distance matching rate of less than 80\%.  The gray diamond at $(R_a,\Delta v_a) \approx (7.5$~pc, 3~\kms) lies at the intersection of the marginalized 50\sth percentile levels for each parameter (and so is reasonably matched to GRS cloud properties) and has a 90\% distance matching rate, but is useful for less than 400 BGPS objects.


\section{RESULTS}\label{ch3:results}

\begin{figure*}[!t]
        \centering
        \includegraphics[width=6.5in]{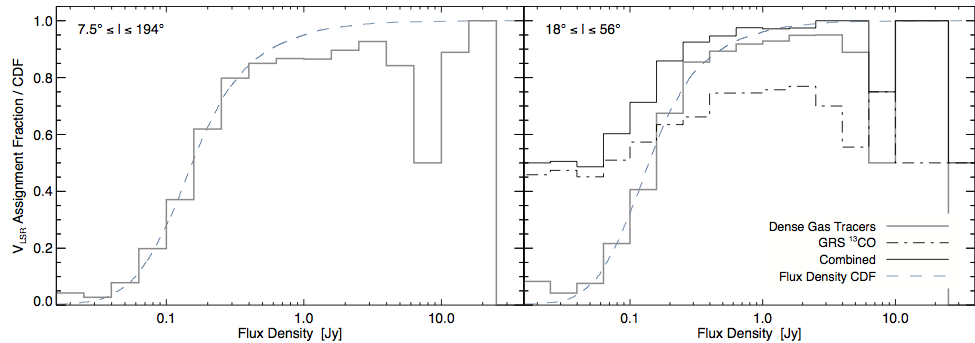}
        \caption[Assignment of \vlsr\ as a function of $\lambda = 1.1$~mm peak flux density.]{Assignment of \vlsr\ as a function of $\lambda = 1.1$~mm peak flux density.  The blue dashed curve in each panel represents the cumulative distribution of flux densities for that subset of sources.  \emph{Left}: Velocity assignment fraction for the collection of dense-gas tracers discussed in \S\ref{data:spec} over the longitude range of \citet{Shirley:2013}.  The dips in assignment fraction at $S_{1.1} \gtrsim 4$~Jy are due to multiple detected velocity components, new V2 sources on the edge of mosaic panels, or source boundary shifts between the V1 and V2.1 catalogs.  \emph{Right}: Galactic longitude further limited to the GRS coverage.  Velocity assignment fractions of dense-gas tracers (solid gray) and morphologically extracted \thco\ spectra (dot-dashed black) are shown, along with the combined velocity assignment fraction (solid black).}
        \label{fig:kin_detect}
\end{figure*}
\subsection{Kinematics}\label{res:kine}

\subsubsection{Dense Gas Velocity Catalogs and BGPS Version 2}

\begin{deluxetable}{lcc}
  \tablecolumns{3}
  \tablewidth{2.0in}
  \tabletypesize{\footnotesize}
  \tablecaption{Number of BGPS V2 Sources for Each Velocity Tracer\label{table:ntracer}}
  \tablehead{
    \colhead{Species}  & \colhead{$N$} & \colhead{Ref.}
  }
  \startdata
  \hcop(3-2) & 2604 & 1 \\
  \nnhp(3-2) & 69 & 1 \\
  CS(2-1) & 256 & 2 \\
  \nhhh(1,1) & 453 & 3,4 \\
  C$^{18}$O(2-1) & 141 & 5 \\
  \thco(1-0) & 2279 & 6
  \enddata
  \tablerefs{1: \citet{Shirley:2013}; 2. Y. Shirley (2012, private communication), 3. \citet{Dunham:2011c}, 4. \citet{Dunham:2010}, 5. M. Lichtenberger (2014, private communication), 6. GRS (this work).}
\end{deluxetable}

The recent re-reduction and expansion of the BGPS data set (G13) has implications for the association of spectroscopic observations with the latest (V2.1)\footnote{The V2.1 catalog corrects various cataloging errors and represents the definitive catalog of objects in the V2.0 images (G13).} continuum source catalog.  BGPS-led spectroscopic surveys were conducted with earlier versions of Bolocat (either 0.7 or 1.0; see S13), and source positions and boundaries may have changed.  These changes generally occur in crowded regions where decomposition is ambiguous or at low signal-to-noise (see G13 for a full discussion).  It is therefore necessary to carefully associate extant \vlsr\ information with the V2.1 source catalog.

The Bolocat cataloging routine produces label maps that identify survey mosaic pixels belonging to each catalog entry \citep[][]{Rosolowsky:2010}.  To account for the finite solid angle encompassed by each spectroscopic pointing, any one \vlsr\ measurement must be associated with all catalog objects whose label mask lies within one beam of the center of the spectroscopic pointing.  For the majority of sources, there is a one-to-one correspondence; however, one velocity pointing may be assigned to more than one catalog source in more crowded regions.  In addition, some pointings are no longer associated with a BGPS V2 source and are ignored.  Use of survey label maps also allows for the direct incorporation of spectroscopic observations not predicated on BGPS catalog positions \citep[\eg observations based on ATLASGAL or Hi-GAL sources;][]{Wienen:2012,Jackson:2013}.

\begin{figure*}[!t]
  \centering
  \includegraphics[width=6.5in]{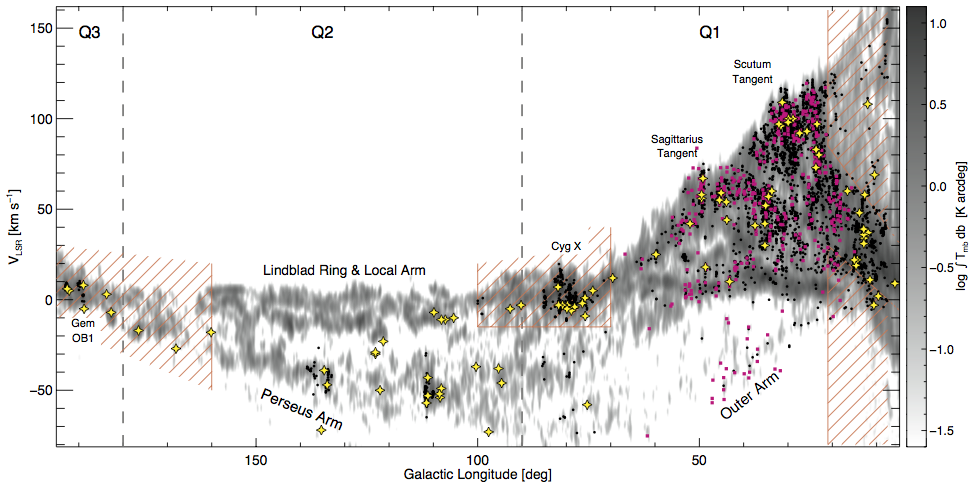}
  \caption[Longitude-velocity diagram for the BGPS.]{Longitude-velocity diagram for the BGPS.  Background image is the latitude-integrated \twco\ intensity of \citet{Dame:2001}.  Black circles mark the locations of BGPS V2 sources, and magenta squares identify the \HII\ regions from Table~\ref{table:hrds}.  Yellow stars mark the trigonometric parallax measurements presented in Table~1 of \citet{Reid:2014}.  The kinematic avoidance zones discussed in \S\ref{res:kaz} are shown as hashed regions.  Various Galactic features are identified.}
  \label{fig:bgps_lv}
\end{figure*}

If more than one spectroscopic pointing is assigned to a given BGPS source, the properties of those spectra are compared.  First, multiple \vlsr\ of spectra within the same survey (\ie \hcop(3-2)) are compared.  If the constituent velocities are more than 5~\kms\ discrepant, no \vlsr\ from that survey is assigned and a flag is returned, otherwise the spectral fit properties (\vlsr, linewidth, uncertainties) are combined, weighted by the peak temperature of each spectrum.  Second, the fit properties of different species are compared (\ie \nhhh(1,1) vs. \hcop(3-2)), and discrepant velocities result in no \vlsr\ being assigned to that source.  Fit properties of different species are combined weighted by signal-to-noise ratio.  

A tabulation of the number of BGPS V2.1 catalog objects which have a valid \vlsr\ from each of the molecular line surveys used is presented in Table~\ref{table:ntracer}.  As an example of the shift in kinematic information from version 1 of the BGPS catalog, the HHT surveys detected either \hcop(3-2) or \nnhp(3-2) emission associated with 3,126 BGPS version 1 sources, but only 2,676 version 2 sources have a valid (\ie not conflicting) \vlsr\ from these data.  For the \nhhh\ surveys of \citeauthor{Dunham:2010}, the number of associated sources went from 490 for V1 to 455 for V2.  There is overlap between the dense-gas surveys, with some catalog objects having accordant velocity information from two or three different molecular line transitions.  Therefore, although the values in Table~\ref{table:ntracer} add up to a larger number, a total of 2,925 BGPS V2 sources have a valid \vlsr\ from one or more of the dense gas spectral surveys discussed in \S\ref{data:spec}.

\subsubsection{Kinematic Catalog Expansion Using \thco}

Because \thco(1-0) traces lower-density gas, the resulting kinematic distance likelihood sometimes represents a different velocity than one of the dense-gas tracers (see Fig.~\ref{fig:13co_hcop}).  The molecular species described in \S \ref{data:spec} trace the dense environments of molecular cloud clumps and cores \citep[][]{Evans:1999}, so we preferentially use that information over the \thco\ data when it is available.  New \thco\ kinematic distances are therefore limited to those objects without an assigned \vlsr\ from the dense-gas tracers due to non-detection, detection of multiple velocity components with $\Delta v \geq 5$~\kms, or sources not observed by spectroscopic surveys.  As a result, while 2,279 Bolocat objects have valid \thco\ \vlsr\ (as defined by the flow chart of Figure~\ref{fig:grs_flow}) only 958 have a \thco\ velocity exclusively.  The remainder overlap one or more dense gas tracer, and have velocities that agree with the other measurements 95\% of the time.  Combination of all available kinematic information yields a collection of 3,900 BGPS sources (representing 45\% of the entire V2 catalog) whose velocity information is included with the expanded distance catalog (\S\ref{res:v2}).

\citet{Shirley:2013} showed a strong correlation between BGPS $\lambda = 1.1$~mm flux density and detection fraction in \hcop(3-2) or \nnhp(3-2).  With the shift to BGPS V2 and the addition of \thco(1-0) as an additional velocity tracer, we reexamine this relationship.  The fraction of sources in logarithmic flux density bins that have an assigned \vlsr\ are shown in Figure~\ref{fig:kin_detect}; sources without such an assignment may be spectroscopic non-detections, have multiple velocity components, or be unobserved.  The left panel of Figure~\ref{fig:kin_detect} reflects the Galactic longitude range of the \citeauthor{Shirley:2013} survey, the most comprehensive sample to date.  The dips in the \vlsr\ assignment rate at $S_{1.1} \geq 4$~Jy are due to 11 catalog sources, 6 of which have multiple velocity components detected in \hcop(3-2), two are new sources at the edge of BGPS mosaics not identified in the V1 catalog, and three are new catalog entries created by the subdivision of V1 sources (see G13 for a discussion) that do not lie within one beam of a spectroscopic pointing.  Aside from these sources, the S13 result that \vlsr\ assignment falls below 50\% for $S_{1.1} < 200$~mJy still holds.

The right panel of Figure~\ref{fig:kin_detect} illustrates this comparison restricted to the Galactic longitude coverage of the GRS.  In addition to the dense-gas tracers (solid gray), the velocity assignment based on morphologically-extracted \thco\ spectra (dot-dashed black) is shown, along with the combined (\ie sources having a velocity from one or the other) assignment rate (solid black).  Two features bear remark: (1) the velocity-assignment rate in \thco\ never falls below 45\% even at low flux density, and (2) the combined \vlsr\ assignment rate $\gtrsim 95\%$ for $S_{1.1} \geq 0.4$~Jy.  The kinematic distance likelihood provided by \thco\ is, therefore, a vital complement to the suite of dense-gas spectroscopic surveys.

\subsubsection{Kinematics and Galactic Structure}\label{res:lv}

The expanded BGPS kinematic catalog can be used to trace out prominent Galactic structure features in the first and second quadrants.  Figure~\ref{fig:bgps_lv} illustrates the correlation between the more diffuse molecular component of the interstellar medium \citep[latitude-integrated \twco\ of][grayscale background]{Dame:2001} and the dense gas seen in the BGPS continuum images (black circles).  The molecular cloud structures identified by the BGPS strongly trace out the Scutum-Centarus arm (whose tangent is at $\ell \approx 30\degr$) and the Sagittarius arm (tangent near $\ell = 50\degr$).  Also prominent are the Cygnus~X star-forming complex near $\ell = 80\degr$ and portions of the Perseus arm from $\ell = 60\degr$ all the way around to Gemini OB1 in the 3\srd quadrant.  BGPS sources at negative \vlsr\ in the 1\sst quadrant are generally associated with CO emission identified as the Outer Arm by \citet{Dame:2001}.  The one feature in the \twco\ image not traced by dense gas is the plethora of molecular gas near \vlsr~= 0~\kms\ in the 2\snd quadrant, identified by \citeauthor{Dame:2001} as being in the Lindblad ring and Local arm.

Locations of HMSFRs, identified by HRDS \HII\ regions or the maser emission targeted by the BeSSeL survey, are also indicated in Figure~\ref{fig:bgps_lv} and generally follow the same patterns as the BGPS molecular cloud structures.  The \HII\ regions from Table~\ref{table:hrds} are shown as magenta squares, and yellow stars mark the BeSSeL maser sources.  While the HRDS only covered the inner Galaxy, its sources pick out the major spiral features including the Outer arm and Perseus arm in the range $60\degr \geq \ell \geq 40\degr$.  The trigonometric parallax measurements of BeSSeL span the entire range of the BGPS kinematic catalog, and provide valuable heliocentric distance anchor points for objects in the hashed kinematic avoidance zones around the Galactic cardinal directions (\S\ref{res:kaz}).

\subsection{Prior DPDFs}\label{res:priors}

\subsubsection{Priors from EB13}

Two prior DPDFs were introduced in EB13 to resolve the KDA for BGPS sources.  The first is based on the distribution of molecular gas in the Galactic disk to constrain sources at high Galactic latitude ($|b| \gtrsim 0\fdg4$) to the near kinematic distance.  This constraint derives from the relatively narrow thickness of the disk's molecular layer \citep[half-width at half maximum $\approx 60$~pc;][]{Bronfman:1988}; lines of sight at these latitudes exit this layer before reaching the far kinematic distance for much of the inner Galaxy.  We introduce here two limitations on the use of DPDF$_\mathrm{H_2}$ that were not relevant to the source sample in EB13.  The model presented by \citet{Bronfman:1988} sought to quantify molecular gas in the ``5-kpc Ring'' and outward through Galactic disk for $R_\mathrm{gal} \geq 0.2 R_0$.  No component is included to model the Galactic central molecular zone.  Additionally, in the outer Galaxy, there is no near/far kinematic discrimination required, and since the density of molecular gas rapidly diminishes as a function of distance in these regions, application of DPDF$_\mathrm{H_2}$ only skews the posterior DPDF to smaller \dsun.  We therefore limit the application of this prior to $11\degr \leq |\ell| \leq 70\degr$, with the upper limit corresponding to the start of the kinematic avoidance zone near $\ell = 90\degr$.

The second prior from EB13 is based on the absorption of mid-infrared light by the dust seen in millimeter continuum surveys.  Expanding on the concept of an infrared dark cloud \citep[IRDC; \cf][]{Simon:2006}, eight-micron absorption features (EMAFs) are visible as decrements in bright Galactic emission in the $\lambda = 8$~\micron\ \spitzer/GLIMPSE images \citep[][]{Churchwell:2009} at the locations of millimeter-detected molecular cloud structures.  By comparing the column density derived from BGPS emission, EB13 placed each object within a numerical model of Galactic 8-\micron\ emission \citep[][]{Robitaille:2012} to construct a DPDF$_\mathrm{emaf}$ based on the morphological comparison of synthetic infrared images with processed versions of the GLIMPSE mosaics.  Application of the EMAF method in the present work remains virtually unchanged from EB13, but with the shift to BGPS V2 and expansion of the kinematic catalog, the collection of sources meeting the automated EMAF selection criteria were once again examined by eye to create the final list of rejected sources (see EB13 for details).  Whereas there were 770 BGPS V1 sources for which DPDF$_\mathrm{emaf}$ was computed, 854 BGPS V2 sources met the final selection criteria, due in part to the expansion of the kinematic data set with GRS \thco\ data.  Of this set of EMAFs, well-constrained distance estimates exist for 679 (see \S\ref{res:v2}), an improvement in itself over the V1 set of 618 well-constrained sources from EB13.

\subsubsection{Trigonometric Parallax Measurements}\label{res:maser}

Using the recently compiled list of 103 trigonometric parallax measurements from the BeSSeL Survey and VERA Project \citep{Reid:2014}, we find a total of 292 BGPS V2 objects that can be associated with one or more maser sources within the confines of the association volume defined above.  Due to the small uncertainty in the parallaxes themselves, 291 of these objects have well-constrained posterior DPDFs (see \S\ref{res:wc}).  The sole unconstrained BGPS source, G023.743-00.235, is associated with a maser identified by \citet{Reid:2014} as being in the 4-kpc / Norma arm, and has a kinematic distance $\sim 1.5$~kpc discrepant from the parallax distance.  Because DPDF$_\mathrm{px}$ is offset from a kinematic distance peak, there still exists sizable probability in both kinematic peaks in the posterior DPDF.  Otherwise, of this sample of BGPS V2 objects, only 223 have a $\mathcal{L}(v_\mathrm{LSR},l,b;d_\sun)$ despite all having a measured \vlsr; the 69 sources lying in a kinematic avoidance zone therefore have well-constrained distance estimates owing exclusively to their association with a trigonometric parallax measurement.

\subsubsection{\HII\ Regions}

The combined HRDS catalog with robust KDA resolutions from Table~\ref{table:hrds} encompasses 441 \HII\ regions, and a total of 525 BGPS V2 sources are associated with one or more of these regions.  The plentiful nature of these objects (see Figure~\ref{fig:bgps_lv}) means that, given the association volume defined in \S\ref{res:assoc}, a BGPS V2 source will occasionally ($\approx 14\%$) be associated with more than one HRDS object.  Approximately 20\% of this subset have an unconstrained DPDF$_\mathrm{hrds}$ based on the conflict between KDA resolutions for constituent \HII\ regions, accounting for some 3\% of the the HRDS-associated BGPS sources.  Overall, 95\% of the HRDS-based DPDFs produce well-constrained distance estimates, with the remaining 2\% being unconstrained sources arising from a disagreement between DPDF$_\mathrm{hrds}$ and DPDF$_\mathrm{emaf}$.

\subsubsection{Gemini OB1}\label{res:gemob1}

\begin{figure}[!t]
        \centering
        \includegraphics[width=3.1in]{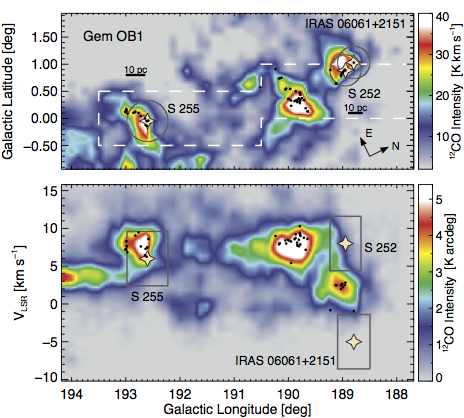}
        \caption[The Gemini OB1 field.]{The Gemini OB1 field.  \emph{Top}: Sky-projected map of the Gemini OB1 molecular cloud.  The background image is the velocity-integrated \twco\ data of \citet{Stacy:1991,Dame:2001}, and black circles mark the locations of BGPS V2 sources.  Stars mark the locations of the masers with trigonometric parallax measurements, and the enclosing circles mark the projected association volumes, computed as in \S\ref{prior:define}.  The white dashed line marks the approximate boundary of the BGPS observations, and the scale bars indicate the projected distances for the two subregions.  \emph{Bottom}: The longitude-velocity integration over $-1\fdg0 \leq b \leq 1\fdg5$ of the same field.  Both panels share a common longitude axis.}
        \label{fig:gem_ob1}
\end{figure}

In the outer Galaxy, star-forming regions tend to be in widely separated, easily distinguished, and coherent structures \citep[][]{Heyer:1998,Brunt:2003,Dunham:2010}.  The Gemini OB1 molecular cloud was one such region selected for detailed study by the BGPS \citep{Dunham:2010}.  Because it lies within a kinematic avoidance zone, its distance must be estimated without $\mathcal{L}(v_\mathrm{LSR},l,b;d_\sun)$.  Fortunately, there exist VLBI trigonometric parallax measurements of \water\ and \methanol\ masers for three sources in this molecular cloud (shown as stars in Figure~\ref{fig:gem_ob1}).  Two of these, S252 and IRAS 06061+2151, lie in the northernmost clump of gas, and have parallax distances of 2.10$^{+0.03}_{-0.03}$~kpc \citep{Reid:2009b} and 2.02$^{+0.53}_{-0.35}$~kpc \citep{Niinuma:2011}, respectively.  The parallax measurement of the \HII\ region S255 (in the more southerly clump of gas) places it somewhat closer at 1.59$^{+0.07}_{-0.07}$~kpc \citep{Rygl:2010a}.  The projections of the cylindrical association volumes for these sources are shown in both panels of Figure~\ref{fig:gem_ob1}.

For the group around S255, most of the BGPS sources lie within the association volume of the parallax measurement, but a handful fall just outside.  Based on the coherent nature of the \twco(1-0) emission in Figure~\ref{fig:gem_ob1}, we assign the 1.59~kpc distance of S255 to the remainder of the BGPS sources in the southerly group.  Interestingly, while the maser locations for S252 and IRAS 06061+2151 are spatially coincident with the northernmost clump of gas and collection of BGPS sources, their velocities bracket the \twco\ and dense-gas emission associated with the dust (Figure~\ref{fig:gem_ob1}, \emph{bottom}).  The origins of these offsets are unclear, and it may be that these masers are associated with young stars that have been ejected from the northerly group of sources.  The limited association volume for applying the parallax prior directly means that none of the BGPS sources in the more northerly group may be assigned a DPDF$_\mathrm{px}$.  Following \citet{Dunham:2010}, therefore, we simply assign the 2.10~kpc distance of S252 to all of the BGPS sources in this complex at $\ell \lesssim 191\degr$.

\subsection{Analysis of ``Well-Constrained'' Distance Estimates}\label{res:wc}

\begin{figure}[!t]
  \centering
  \includegraphics[width=3.1in]{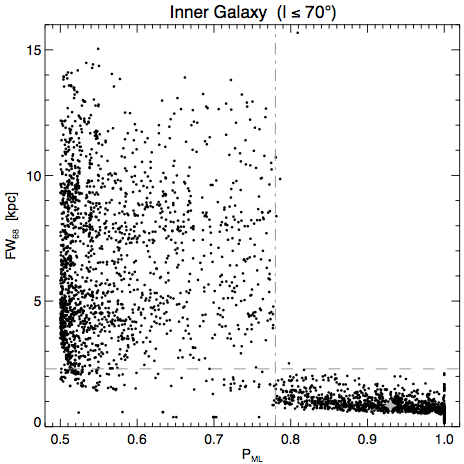}
  \caption[``Well-constrained'' distance estimate criteria.]{``Well-constrained'' distance estimate criteria.  Each BGPS V2.1 catalog object at $\ell \leq 70\degr$ is shown, plotted by its integrated posterior DPDF on the \dml\ side of the tangent point (\pchoose) and the full-width error bar containing $\geq 68.3\%$ of the integrated posterior DPDF around \dml\ (FW$_{68}$).  The vertical dot-dashed line at \pchoose~=~0.78 is shown to guide the eye, while the horizontal dashed line marks the FW$_{68} \leq$ 2.3~kpc boundary for a ``well-constrained'' distance estimate.  The gray star identifies the median values of \pchoose\ and FW$_{68}$ for the well-constrained sample.}
  \label{fig:pml_wc}
\end{figure}

The definition of a ``well-constrained'' distance estimate was characterized in EB13 for objects identified as EMAFs.  With the introduction of additional kinematic distance likelihoods and prior DPDFs, we analyze the properties of the newly-accumulated posterior DPDFs in the event different criteria are required for this definition.  As a proxy for the strength of the KDA resolution, \pchoose\ is the integrated posterior DPDF on the side of the tangent point (\dtan) containing the maximum-likelihood distance (\dml).  For sources well away from \dtan\ this includes the entirety of a single kinematic distance peak.  Because the posterior DPDFs are normalized to unit integral probability and there are at most two possible kinematic distances, $0.5 \lesssim$ \pchoose~$\leq 1.0$.  The measure of constraint tightness on a distance estimate is computed as the full-width of the region around \dml\ containing at least 68.3\% of the integrated probability and whose bounds occur at equal probability level (FW$_{68}$).  Sources near \dtan\ will have FW$_{68}$~$\approx 1-2$~kpc due to localization from $dv_{_\mathrm{LSR}} / d(d_{_\sun})$ (\S\ref{res:kaz}).  For single-peaked DPDFs, FW$_{68}$ approximates the Gaussian $\pm 1\sigma$ region around the centroid, but this approximation breaks down when one kinematic peak is not dominant over the other.

Choice of criteria for ``well-constrained'' distance estimates once again relies on the bivariate distribution of \pchoose\ and FW$_{68}$, shown in Figure~\ref{fig:pml_wc}.  As in EB13, the pattern whereby FW$_{68}$ becomes large for \pchoose~$\lesssim 0.78$ holds.  This break is due to the geometry of $\mathcal{L}(v_\mathrm{LSR},l,b;d_\sun)$, where a single kinematic distance peak may contain $\geq 68.3\%$ of the integrated posterior DPDF only if it sufficiently dominates over the other peak.  For objects that have \pchoose~$\gtrsim 0.78$, the maximum value of FW$_{68}$ is approximately 2.3~kpc.  We again choose to define a ``well-constrained'' distance estimate as FW$_{68}$~$\leq 2.3$~kpc to include objects within $\approx 1$~kpc of \dtan\ (lower-left corner of Figure~\ref{fig:pml_wc}) because the near/far ambiguity for these sources leads to a distance discrepancy of $\lesssim 2$~kpc.  We note that while we chose FW$_{68} = 2.3$~kpc to define ``well-constrained'', it is clearly an upper limit, and the vast majority of these sources have FW$_{68} \lesssim 1.5$~kpc, or a distance uncertainty of $\lesssim 0.8$~kpc.  The median values of \pchoose\ and FW$_{68}$ for the well-constrained sample are 0.928 and 0.84~kpc, respectively (gray star in Figure~\ref{fig:pml_wc}).

In the outer Galaxy, there is neither a tangent distance nor an ambiguity in the kinematic distance, so all distance estimates for objects in Quadrants II and III have \pchoose~$= 1.0$ and  FW$_{68}$ well within the limit outlined above.

\subsection{The V2 Distance Catalog and Posterior DPDFs}\label{res:v2}

\subsubsection{Catalog Description}

\begin{deluxetable}{lcccc}
  \tablecolumns{5}
  \tablewidth{0pt}
  \tabletypesize{\footnotesize}
  \tablecaption{KDA Resolutions for BGPS V2 Sources\label{table:kdars}}
  \tablehead{
    \colhead{KDA} & \colhead{Flag} & \colhead{$N_\mathrm{kin}$\tablenotemark{a}} & \colhead{$N_\mathrm{tot}$\tablenotemark{b}}  & \colhead{$f_\mathrm{w.c.}$\tablenotemark{c}}\\
     \colhead{Resolution} & \colhead{} & \colhead{} & \colhead{}  & \colhead{(\%)}
  }
  \startdata
  Near & N & 984 & 984 & 57.5 \\
  Far & F & 336 & 336 & 19.6 \\
  Tangent & T & 193 & 193 & 11.3 \\
  Outer Galaxy & O & 197 & 197 & 11.5 \\
  Unconstrained & U & 1798 & 6487 & \nodata \\
  Excluded\tablenotemark{d} & X & 0 & 397 & \nodata \\
  \hline
  Total & & 3508 & 8594 & 100
  \enddata
  \tablenotetext{a}{Number of objects in the ``kinematic sample''.}
  \tablenotetext{b}{Number of objects from the full Bolocat V2.}
  \tablenotetext{c}{Fraction of the well-constrained sources with this KDA resolution.}
  \tablenotetext{d}{Object lies in a kinematic avoidance zone (\S\ref{res:kaz}).}
\end{deluxetable}

\begin{deluxetable*}{cllcclccccccclc}
  \tablecolumns{15}
  \tablewidth{0pc}
  \tabletypesize{\footnotesize}
  \tablecaption{BGPS V2 Distance Catalog\label{table:dists}}
  \tablehead{
    \multicolumn{4}{c}{V2.1 Catalog Properties} & \colhead{} & \multicolumn{2}{c}{Velocity} & \colhead{} & \multicolumn{4}{c}{Heliocentric Distance} & \colhead{} & \multicolumn{2}{c}{Galactocentric Position} \\
    \cline{1-4} \cline{6-7} \cline{9-12} \cline{14-15}
    \colhead{Catalog}  & \colhead{$\ell$} & \colhead{$b$} & \colhead{$S_\mathrm{int}$} & \colhead{} & \colhead{\vlsr} & \colhead{Ref.} & \colhead{} & 
\colhead{KDA} & \colhead{\pchoose\tablenotemark{a}} & \colhead{\dml} & \colhead{\dbar} & \colhead{} & \colhead{$R_\mathrm{gal}$\tablenotemark{b}} & \colhead{$z$\tablenotemark{c}} \\
    \colhead{Number} & \colhead{(\degr)} & \colhead{(\degr)} & \colhead{(Jy)} & \colhead{} & \colhead{(\kms)} & \colhead{} & \colhead{} & 
\colhead{Resol.} & \colhead{} & \colhead{(kpc)} & \colhead{(kpc)} & \colhead{} & \colhead{(kpc)} & \colhead{(pc)}
  }
  \startdata
3475 & \phn20.441 & $-0.030$ & $0.72(0.17)$ & & \phs\phn$76.0$ & 3 & & F & 0.99 & $11.00_{-0.24}^{+0.26}$ & \nodata & & \phn$4.32_{-0.17}^{+0.20}$ & \phd\phn$-12(5.1)$ \\
4061 & \phn24.171 & \phs$0.063$ & $0.56(0.11)$ & & \phs$110.4$ & 6  & & N & 0.78 & \phn$5.60_{-0.42}^{+0.44}$ & \nodata & & \phn$3.96_{-0.20}^{+0.23}$ & \phs\phd\phn$16(5.0)$ \\
4969 & \phn29.279 & \phs$0.096$ & $1.39(0.14)$ & & \phs\phn$87.0$ & 5 & & N & 0.79 & \phn$4.82_{-0.56}^{+0.58}$ & \nodata & & \phn$4.76_{-0.27}^{+0.31}$ & \phs\phd\phn$20(5.0)$ \\
5197 & \phn30.464 & \phs$0.033$ & $1.60(0.18)$ & & \phs$106.0$ & 1,2 & & U & 0.51 & \nodata & \nodata & & \phn$4.46^{+0.14}_{-0.14}$ & \nodata \\
5560 & \phn31.715 & \phs$0.502$ & $0.31(0.09)$ & & \phs\phn$53.6$ & 6  & & U & 0.65 & \nodata & \nodata & & \phn$5.85^{+0.23}_{-0.23}$ & \nodata \\
6196 & \phn35.043 & $-0.474$ & $5.14(0.34)$ & & \phs\phn$51.2$ & 1 & & F & 0.79 & $10.38_{-0.58}^{+0.60}$ & \nodata & & \phn$5.96_{-0.33}^{+0.38}$ & \phd\phn$-86(8.1)$ \\
6711 & \phn43.816 & $-0.120$ & $0.39(0.14)$ & & \phs\phn$46.0$ & 6  & & T & 0.65 & \phn$6.08_{-0.20}^{+0.18}$ & $6.09(0.20)$& & \phn$5.77_{-0.00}^{+0.01}$ & \phn\phn$-0.9(5.1)$ \\
6964 & \phn53.591 & $-0.244$ & $1.31(0.24)$ & & \phs\phn$59.7$ & 1,3 & & T & 0.54 & \phn$4.86_{-0.78}^{+0.78}$ & $4.86(0.77)$& & \phn$6.71_{-0.00}^{+0.05}$ & \phn\phn$-4.4(6.8)$ \\
7612 & 110.047 & $-0.019$ & $0.14(0.05)$ & & \phn$-50.7$ & 1 & & O & 1.00 & \phn$4.32_{-0.62}^{+0.68}$ & \nodata & & $10.63_{-0.41}^{+0.47}$ & \phs\phd\phn$28(5.0)$ \\
8210 & 192.584 & $-0.043$ & $13.93(0.95)$ & & \phs\phn\phn$8.6$ & 1,4 & & O & 1.00 & \phn$1.60_{-0.06}^{+0.06}$ & \nodata & & \phn$9.91_{-0.06}^{+0.06}$ & \phs\phd\phn$28(5.0)$ 
  \enddata
  \tablenotetext{a}{The integrated posterior DPDF on the \dml\ side of the tangent point; used as a proxy for the quality of the distance constraint.}
  \tablenotetext{b}{$R_\mathrm{gal}$ is unambiguous and is computed from \vlsr\ and assumed uncertainty of 7~\kms\ \citep{Reid:2009} for sources without a well-constrained distance estimate.}
  \tablenotetext{c}{Errors include contributions from variations in $z$ along the line of sight over the range $d_{_\sun} \pm \sigma_d$ and the $\pm 5$~\kms\ uncertainty in the solar offset above the Galactic midplane \citep{Juric:2008}, added in quadrature.}
  \tablerefs{1: \citet{Shirley:2013}; 2. Y. Shirley (2012, private communication), 3. \citet{Dunham:2011c}, 4. \citet{Dunham:2010}, 5. M. Lichtenberger (2014, private communication), 6. GRS (this work).}  
  \tablecomments{Errors are given in parentheses.}
  \tablecomments{This table is available in its entirety in a machine-readable format in the online journal.  A portion is shown here for guidance regarding its form and content.}
\end{deluxetable*}

A posterior DPDF for each BGPS V2.1 catalog object was computed using Equation~(\ref{eqn:ch3_dpdf}) and the appropriate kinematic distance likelihood (dense gas or GRS \thco), and all applicable prior DPDFs.  From the posterior DPDFs, the \pchoose\ and FW$_{68}$ statistics, distance estimates and KDA resolutions were determined.  The resulting distance catalog includes relevant information from the Bolocat V2.1, velocity information (\vlsr, survey), and heliocentric distance and Galactocentric position, if available.

With the expansion of the source list beyond $\ell = 65\degr$, two new KDA resolution flags are introduced in this distance catalog.  As in EB13, sources whose \dml\ is within 1~kpc of \dtan\ are given the flag~T, indicating they are at (or near enough) the tangent point.  Sources in the inner Galaxy with ``well-constrained'' distance estimates with \dml~$< (>)$ \dtan\ are again given the flag~N (F).  Outer-Galaxy objects, for which there is no KDA, are assigned O if they have an associated \vlsr\ and lie outside a kinematic avoidance zone.  Those objects throughout the Galactic plane lying inside a kinematic avoidance zone are given the flag~X, specifying their kinematic information has been excluded.  Objects in a kinematic avoidance zone that are associated with a trigonometric parallax measurement or are in Gemini OB1 may still be given a ``resolved'' KDA flag.  The remaining sources which either have no kinematic information or whose posterior DPDF does not meet the criteria of \S\ref{res:wc} are assigned the flag~U.  Table~\ref{table:kdars} lists the number of objects in the BGPS catalog with each KDA resolution flag.  Column 3 lists objects in the ``kinematic sample'', that is objects that possess either a kinematic distance from dense gas or \thco\ spectra, an association with a trigonometric parallax measurement, or a location in Gemini OB1.  The fourth column lists all BGPS V2 objects, while the final column indicates the fraction of well-constrained sources with each KDA resolution flag.

The BGPS distance catalog is presented in Table~\ref{table:dists}, which contains entries for each of the 3,689 catalog objects in the kinematic sample.  Objects with a well-constrained distance estimate (flags N/F/T/O) have the maximum-likelihood distance (\dml) listed, along with the associated error bars.  Tangent point objects additionally list the first-moment distance (\dbar), following the discussion in EB13.  Object with flags U or X have no heliocentric distance information included.  Galactocentric radius is computed for each object with a detected \vlsr, save those with KDA flag~X, as $R_\mathrm{gal}$ is not subject to the KDA but is affected by the non-circular motions characterizing the kinematic avoidance zones.  For objects with a well-constrained distance estimate, Galactocentric vertical position ($z$) is also computed, subject to the coordinate transformation presented in Appendix C of EB13.  The DPDFs for all 8,594 objects in the BGPS V2.1 catalog are publicly available.\footnote{Available through IPAC at\\ \texttt{http://irsa.ipac.caltech.edu/data/BOLOCAM\_GPS}}

\begin{figure*}[!t]
  \centering
  \includegraphics[width=6.5in]{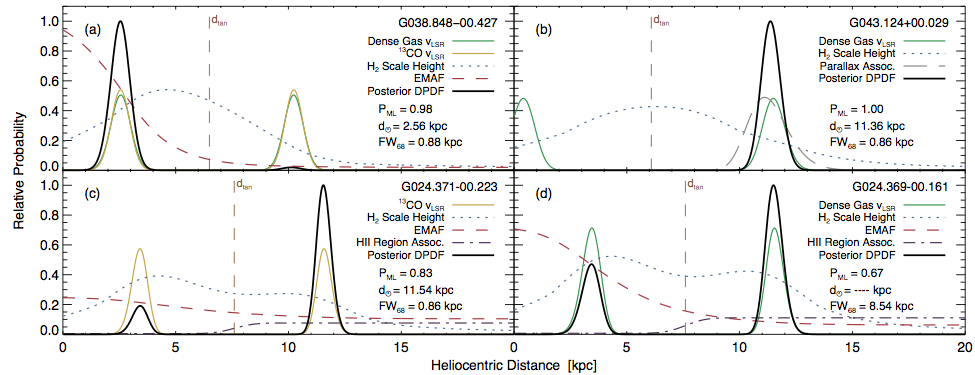}
  \caption[Collected DPDFs for example sources.]{Collected DPDFs for example sources.  (\emph{a}) Source G038.848-00.427 has a \thco\ velocity agreeing with the dense gas \vlsr, and is placed at the near kinematic distance by DPDF$_\mathrm{emaf}$ with high probability.  (\emph{b}) Source G043.124+00.029 is associated with the trigonometric parallax measurement of W49 N.  (\emph{c}) Source G024.371-00.223 has kinematic information from \thco\ and also has conflicting DPDF$_\mathrm{emaf}$ and DPDF$_\mathrm{hrds}$, but the latter contributes more strongly to marginally place this source at the far kinematic distance.  (\emph{d}) Source G024.389-00.161 is associated with the same \HII\ region as the source in \emph{c}, but the stronger contribution from DPDF$_\mathrm{emaf}$ renders the distance to this source ambiguous.}
  \label{fig:dpdf_example}
\end{figure*}

DPDFs for example sources are shown in Figure~\ref{fig:dpdf_example} for several cases.  First, the constituent DPDFs provide a well-constrained near distance for source G038.848-00.427 (panel \emph{a}), with the DPDF$_\mathrm{emaf}$ providing most of the resolving power.  This source also demonstrates a \thco\ velocity that agrees with the \vlsr\ from a dense gas tracer.  Panel (\emph{b}) illustrates a source (G043.124+00.029) associated with a trigonometric parallax measurement, W49 in this case.  The DPDF$_\mathrm{px}$ aligns well with the kinematic distance likelihood to produce a tightly-constrained distance.  An example of a marginal distance resolution is shown in panel (\emph{c}), where source G024.371-00.223 has conflicting DPDF$_\mathrm{emaf}$ and DPDF$_\mathrm{hrds}$.  In this case, the HRDS prior contributes more strongly to the posterior DPDF, placing enough probability in the far kinematic distance peak to constrain the distance.  A neighboring source (G024.369-00.161, panel \emph{d}), however, has a stronger EMAF morphological match that leaves the posterior DPDF without sufficient probability in a single peak for a constrained distance.

\subsubsection{Source Properties}\label{res:props}

\begin{figure*}[!t]
        \centering
        \includegraphics[width=3.1in]{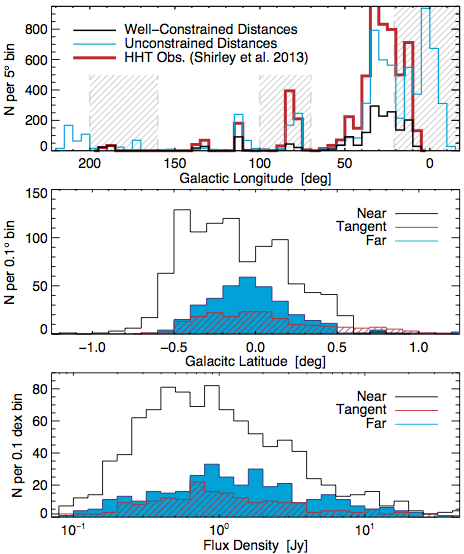}
        \includegraphics[width=3.1in]{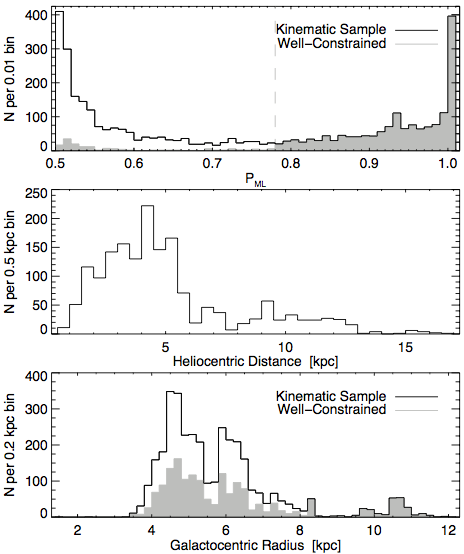}
        \caption[Summary of source properties from Table~\ref{table:dists}.]{Summary of source properties from Table~\ref{table:dists}.  \emph{Top Left}: Comparison of the Galactic longitude distributions for objects with well-constrained (black) versus unconstrained (cyan) distances, with the red histogram showing the distribution of spectroscopic observations of \citet{Shirley:2013}.  The gray hashed regions mark the longitude-projected kinematic avoidance zones (\S\ref{res:kaz}).  \emph{Middle Left}: Distributions of Galactic latitude for sources with well-constrained distance estimates at $\ell \leq 90\degr$.  Colors represent near (black), far (blue), and tangent (red) KDA resolutions.  \emph{Bottom Left}: As above, but showing the distributions of $\lambda = 1.1$~mm flux density.  \emph{Top Right}: Distributions of \pchoose\ for the entire kinematic sample (black) and sources with well-constrained distance estimates (gray).  \emph{Middle Right}: Heliocentric distance distribution for the well-constrained subset.  \emph{Bottom Right}: Galactocentric radius distributions for the entire kinematic sample (black) and sources with well-constrained distance estimates (gray).}
        \label{fig:glon}\label{fig:source_hist}
\end{figure*}

From the pool of sources in Table~\ref{table:dists}, 1,710 (49\%) have a well-constrained distance estimate, representing a substantial population of molecular cloud structures for which physical properties may be derived (T. Ellsworth-Bowers, in preparation).  A summary of the quantities in Table~\ref{table:dists} is presented in Figure~\ref{fig:glon} and discussed here.

The distribution of Galactic longitude (top-left panel) picks out concentrations of sources along the Galactic plane.  The black histogram represents the set of well-constrained distances, while cyan represents the remainder of the BGPS V2.1 catalog.  Gray hashing delineates the longitude-projected kinematic avoidance zones (\S\ref{res:kaz}), although only the region around $\ell = 180\degr$ applies to all \vlsr.  Any well-constrained objects within these zones either lie at an allowed \vlsr\ or are associated with a trigonometric parallax measurement (\S\ref{prior:maser}).  The red histogram describes the distribution of spectroscopic observations from S13, illustrating the regions for which distances could possibly be derived (\eg S13 did not observe $\ell \geq 200\degr$, as this was a new region in the BGPS V2 release; G13).

The histograms of Galactic latitude and $\lambda = 1.1$~mm flux density are shown for sources at $\ell \leq 90\degr$ in the middle- and bottom-left panels of Figure~\ref{fig:source_hist}, respectively, for each of the three KDA resolutions to illustrate the systematic effects of sources at different distances.  Outer Galaxy sources are excluded to minimize the effects of Galactocentric radius on the analysis.  In the middle-left panel, objects at the near kinematic distance or tangent point subtend larger swaths across the width of the Galactic plane (FWHM~$\approx 0\fdg8$ and $\approx 0\fdg9$, respectively) than the far kinematic group (FWHM~$\approx 0\fdg6$).  This is to be expected as objects at $|b| \gtrsim 0\fdg4$ are generally assigned the near kinematic distance by DPDF$_\mathrm{H_2}$.  A Kolmogorov-Smirnov (K-S) test finds that even the far and tangent groups come from different underlying distributions at a 99.7\% confidence level, while the near group is further divergent.  In terms of the millimeter flux density, the median values for the near and tangent groups are $\approx 0.8 - 0.9$~Jy, and that of the far group is 1.1~Jy.  The slightly greater median for the far group implies that the fainter objects detected by the BGPS tend to be nearby.  A future release of Bolocat will utilize algorithms for separating filamentary versus compact emission, and will be able to investigate whether fainter objects tend to be more filamentary, and therefore not resolvable or detectable at great distance.  A K-S test shows that while the near and far groups are drawn from different distributions (at the 99.998\% confidence level), the tangent group seems to straddle the fence, as it cannot be ruled out that it differs from the near or the far groups at the 95\% confidence level.

The right panels in Figure~\ref{fig:source_hist} illustrate the distance resolution aspects of Table~\ref{table:dists}.  The distribution of \pchoose\ for the kinematic sample and subset of well-constrained sources is plotted in the top-right panel, with the overlap being complete for \pchoose~$\geq 0.78$.  The well-constrained sources with \pchoose~$\approx 0.5$ are near the tangent point, as the peaks in the posterior DPDFs straddle \dtan.  The middle-right panel shows the heliocentric distance distribution of well-constrained sources, with 1237/1710 (72\%) of sources nearer than 5.5~kpc.  Since Galactocentric radius is not subject to the KDA, the bottom-right panel shows the distributions for both the kinematic sample and the well-constrained subset.  There is a strong break at $R_\mathrm{gal} \approx 5.5$~kpc, which may represent the division between the Scutum-Centarus and Sagittarius arms or it may be an artifact of the chosen flat rotation curve.\footnote{\citet{Persic:1996} derived a ``universal'' spiral galaxy rotation curve from the observed rotation curves of over 1,000 galaxies that shows a clear downturn in the circular velocity in the inner several kiloparsecs.  This type of downturn is consistent with the measured parallaxes and proper motions for Galactic HMSFRs of \citet{Reid:2014}, indicating that a flat rotation curve is not valid for $R_\mathrm{gal} \lesssim 5$~kpc.}  Kinematic distances are unambiguous for $R_\mathrm{gal} > R_0$, so all objects in the kinematic sample beyond the solar circle have well-constrained distance estimates.  The marked gap at \rgal~= $8.5-9.5$~kpc is the result of the only spiral feature (Perseus arm) within the BGPS coverage region with appreciable gas in this Galactocentric radius range lying within a kinematic avoidance zone.

\subsection{Galactocentric Positions}\label{res:galaxy}

One important application of a large collection of well-constrained distance estimates for molecular cloud structures is the elucidation of Galactic structure in terms of the dense molecular gas that hosts star formation.  Galactocentric positions may be derived using the ($\ell,b,$\dsun$) \rightarrow (R_\mathrm{gal},\phi,z)$ conversion matrix from Appendix C of EB13, which accounts for the $\approx 25$~pc vertical offset of the Sun above the Galactic midplane \citep{Humphreys:1995,Juric:2008}.

\subsubsection{Face-On View of the Milky Way}\label{res:face}

\begin{figure}[!t]
        \centering
        \includegraphics[width=3.1in]{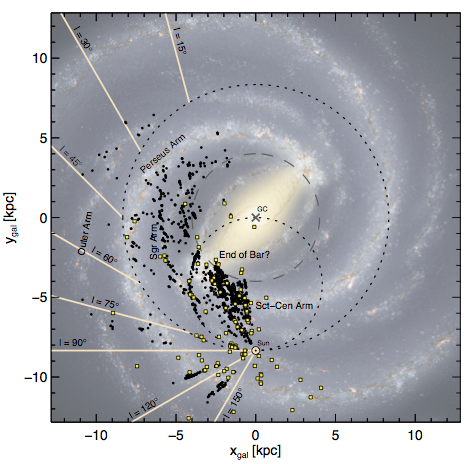}
        \caption[Face-on view of the Milky Way for sources with well-constrained distance estimates, plotted atop an artist's rendering of the Milky Way (R. Hurt: NASA/JPL-Caltech/SSC) viewed from the north Galactic pole.]{Face-on view of the Milky Way for sources with well-constrained distance estimates (black circles), plotted atop an artist's rendering of the Milky Way (R. Hurt: NASA/JPL-Caltech/SSC) viewed from the north Galactic pole.  Yellow squares mark the locations of masers with trigonometric parallaxes \citep[][Table~1]{Reid:2014}.  The image has been scaled to match the $R_0$ used for calculating kinematic distances.  The outer dotted circle marks the solar circle, and the inner dotted circle the tangent point as a function of longitude.  The dashed circle at $R_\mathrm{gal} = 4$~kpc outlines the region influenced by the long Galactic bar where the assumed flat rotation curve breaks down \citep{Benjamin:2005,Reid:2014}.  Various suggested Galactic features are labeled.  For clarity, distance error bars are not shown.  (A color version of this figure is available in the online journal.)}
        \label{fig:mw}
\end{figure}

\begin{figure*}[!t]
        \centering
        \includegraphics[width=2.1in]{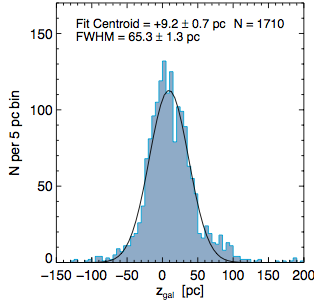}
        \includegraphics[width=2.1in]{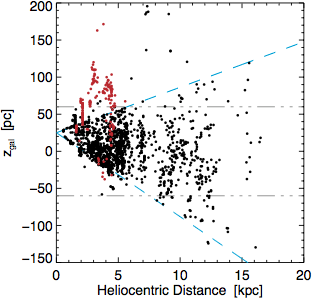}
        \includegraphics[width=2.1in]{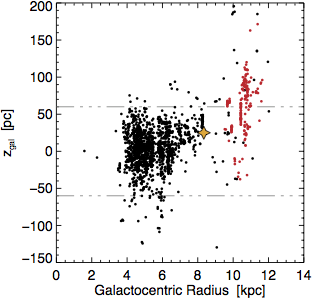}
        \caption[Vertical distribution of sources about the Galactic midplane.]{Vertical distribution of sources about the Galactic midplane.  \emph{Left}: Histogram of $z$ with gaussian fit overplotted.  \emph{Center}: Vertical position as a function of heliocentric distance, with cyan dashed lines showing approximate boundaries of BGPS coverage ($|b| \leq 0\fdg5$) at $\ell = 30\degr$.  Sources plotted in red are at $\ell > 90\degr$.  The gray dot-dashed lines mark the 60-pc scale height of molecular gas \citep{Bronfman:1988}.  \emph{Right}: Vertical position as a function of Galactocentric radius.  Red sources and gray dot-dashed lines as in the middle panel.  The star marks the Sun's location.}
        \label{fig:z}
\end{figure*}

The face-on map of the Milky Way from the north Galactic pole is shown in Figure~\ref{fig:mw}, with the maximum-likelihood distance (or \dbar\ for sources near \dtan) for each well-constrained source plotted atop an illustration of the Galaxy derived from \spitzer\ near-infrared stellar data (R. Hurt: NASA/JPL-Caltech/SSC), scaled to the $R_0$ from Table~\ref{table:mw}.  There are two key attributes of this figure that bear mentioning.  The first is the spread in heliocentric distance of sources along any given line of sight (most noticeable around $\ell = 30\degr$ and $\ell = 110\degr$).  Error bars are not shown in Figure~\ref{fig:mw} for clarity, but with a typical uncertainty of $\approx 0.5$~kpc, object positions within various complexes are self-consistent.  In addition to the uncertainty inherent in the posterior DPDF, deviant motions of the gas away from circular motion around the Galactic center can offset \vlsr\ and shift derived distances away from their true position, thereby creating an apparent dispersion along the line of sight and smoothing the underlying structure.

The second notable attribute is the placement of BGPS objects in regions of the background image that appear devoid of stars in the model (\ie $\ell \approx 30\degr \pm 10\degr$, or $(x_\mathrm{gal},y_\mathrm{gal}) \approx (-4,0)$~kpc).  There are two possible interpretations.  The first is that the background image is a ``best guess'' only, based on stellar distributions from \spitzer\ data.  Robust distance measurements for molecular cloud clumps may well be telling a different story of the locations of spiral arms and Galactic structure.  For example, \citet{Egusa:2011} found a significant population of molecular gas ``downstream'' of spiral arms in M51, nearly spanning the interarm region and coincident with \HII\ regions identified in near-infrared images.  The second interpretation is these are objects incorrectly placed at the far kinematic distance by the set of prior DPDFs currently implemented.  In this case, it is likely that, as the suite of data-driven prior DPDFs grows, sources will shift away from these vacant regions in the \spitzer\ model.  The $\approx 80$ BGPS sources in this area are primarily associated with HRDS \HII\ regions; a (future, undeveloped) self-consistent \HI\ absorption prior DPDF may solve this mystery.

Notwithstanding uncertainties in source location in Figure~\ref{fig:mw}, several prominent Galactic features begin to suggest themselves based on the BGPS V2 distance catalog.  The most significant is the end of the Galactic bar near $\ell = 30\degr$ and the start of the Scutum-Centarus Arm moving to smaller longitude.  Next is the general outline of the Sagittarius arm, visible from ($x_\mathrm{gal},y_\mathrm{gal}) \approx (-3,3)$~kpc counterclockwise around to its tangency near $\ell = 50\degr$.  Portions of the Perseus arm are traceable in the $\ell = 40\degr - 50\degr$ region and again in the outer Galaxy.  Finally, the BGPS detects 23 objects in the Outer arm beyond the solar circle in the $\ell = 20\degr - 80\degr$ range, at a heliocentric distance of $\approx 10-15$~kpc.

\subsubsection{Vertical Distribution of Star Formation}\label{res:vert}

\begin{figure*}[!t]
        \centering
        \includegraphics[width=6.5in]{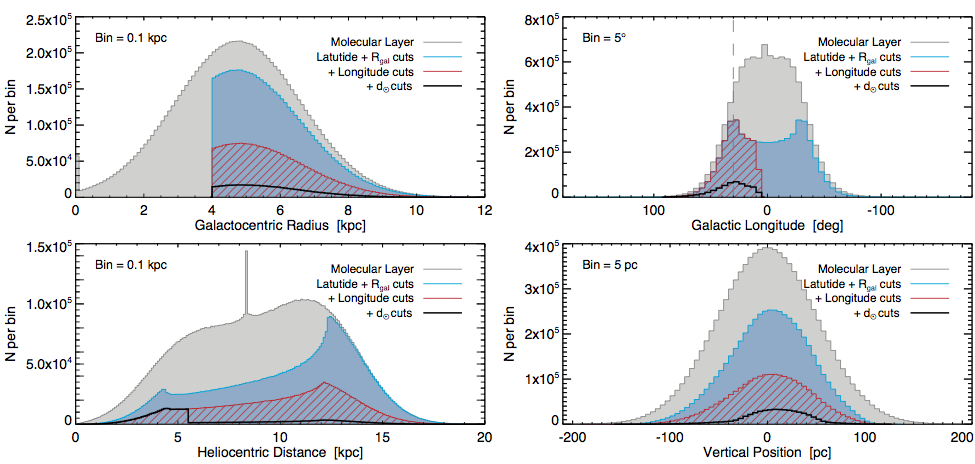}
        \caption[Analysis of the observational effects on the measured vertical scale height of dense molecular cloud structures.]{Analysis of the observational effects on the measured vertical scale height of dense molecular cloud structures.  In each panel, the histograms represent (gray) sources spread throughout the entire molecular layer, (blue) sources meeting the $|b| \leq 0\fdg5$ and \rgal~$\geq 4$~kpc limits of the BGPS, (red) sources additionally meeting the $90\degr \geq \ell \geq 7\fdg5$ longitude limits of the kinematic sample in the inner Galaxy, and (black) sources additionally meeting heliocentric distance criteria to match the observed sample (see text).  The panels show the distributions of sources in Galactocentric radius (top left), Galactic latitude (top right), heliocentric distance (bottom left), and vertical position about the Galactic midplane (bottom right).}
        \label{fig:vert_app}
\end{figure*}

In addition to the face-on map of the Milky Way, well-constrained distance estimates permit study of the vertical distribution ($z$) of sources about the Galactic midplane (Figure~\ref{fig:z}).  The errors tabulated in the last column of Table~\ref{table:dists} include contributions from variations in $z$ along the line of sight over the range $d_{_\sun} \pm \sigma_d$ and the $\pm 5$~\kms\ uncertainty in the solar offset above the Galactic midplane \citep{Juric:2008}, added in quadrature.  The left panel depicts the histogram of $z$, which may be fit by a Gaussian with a centroid at $+9.2\pm0.7$~pc, a FWHM of $65.3\pm1.3$~pc, and a reduced $\chi^2_\mathrm{red} = 1.8$.  The centroid being at slight positive $z$ should not be confused with a centroid at slight positive Galactic latitude.  In the middle panel, however, it is apparent that the width and centroid of the distribution may be slightly misleading owing to the nominal $|b| \leq 0\fdg5$ limit of BGPS coverage.  The cyan dashed lines in that panel mark this limit at $\ell = 30\degr$ (these limits rotate to more positive values at larger longitude owing to the Sun's vertical displacement above the $z=0$ plane).  In both the middle and right panels, red circles mark BGPS sources in the outer Galaxy ($\ell > 90\degr$) where survey coverage was neither blind nor uniform, but rather focused on known regions of star formation.  The gray dot-dashed lines mark the FWHM of the Galactic molecular layer \citep[=~120~pc;][]{Bronfman:1988}.  The BGPS does not probe the full width of the molecular layer until \dsun~$\gtrsim 6$~kpc, whereas the bulk of the distance catalog ($\approx 76\%$) is closer than this point.  The FWHM of the distribution in the left panel, therefore, should be viewed as a lower limit on the scale height of dense star-forming gas in the Galactic plane (see \S\ref{disc:vertical} for a discussion of the observational effects of the BGPS on the derived vertical position distribution).  The rightmost panel in Figure~\ref{fig:z} illustrates the relationship between Galactocentric radius and vertical position; the orange star marks the Sun's location.  Visible here is a warp in the molecular disk beyond the solar circle.


\section{DISCUSSION}\label{ch3:discuss}

\subsection{On the Vertical Distribution and Galactic Census of Dense Molecular Cloud Structures}\label{disc:vertical}

The vertical distribution of dense molecular cloud structures from the present subset of BGPS sources seems to imply that the star-forming structures in the interstellar medium have a smaller scale height than that measured for the molecular gas.  As noted in \S\ref{res:vert}, however, the nominal $|b| \leq 0\fdg5$ limit of the BGPS may impose an observational bias on this result.  Here we investigate the connection between the observed distribution of sources with well-constrained distances and the Galactic distribution of molecular gas.

We conducted a series of simulations whereby objects were randomly distributed according to the axisymmetric \htwo\ distribution of \citet[][derived from the observations of \citealp{Bronfman:1988}]{Wolfire:2003}, then ``observed'' through a series of increasingly restricting criteria.  The (\rgal,$\phi,z$) positions of 40,000 simulated sources were drawn randomly from the distributions of each coordinate from the \citeauthor{Wolfire:2003} model, and then converted to the observational \lbd\ coordinates using the inverse of the matrix operation from Appendix~C of EB13.  This number of objects was chosen so that the final subset was comparable in number to the set of well-constrained distances for BGPS objects presented in this paper.

\begin{deluxetable}{lccc}
  \tablecolumns{4}
  \tablewidth{0pc}
  \tabletypesize{\small}
  \tablecaption{Vertical Distribution Fits for Subsets\label{table:vert_app}}
  \tablehead{
	\colhead{Set / Cut} & \colhead{Centriod} & \colhead{FWHM} & \colhead{$N$} \\
	\colhead{} & \colhead{(pc)} & \colhead{(pc)}	& \colhead{}
  }
  \startdata
Molecular Layer & $\phn0.0 \pm 0.3$ & $120.0 \pm 0.6$ & $40000 \pm \phn0$ \\
Latitude + \rgal\ cuts & $\phn4.1 \pm 0.4$ & $\phn98.8 \pm 0.7$ & $21318 \pm 92$ \\
+ Longitude cuts & $\phn4.3 \pm 0.5$ & $\phn97.4 \pm 1.1$ & $\phn9180 \pm 77$ \\
+ \dsun\ cuts & $10.9 \pm 0.8$ & $\phn67.5 \pm 1.5$ & $\phn1987 \pm 33$
  \enddata
\end{deluxetable}

Since the molecular hydrogen \rgal\ distribution is modeled as a Gaussian with sufficient width to have non-zero probability at \rgal~= 0, any randomly drawn negative value was assigned to the Galactic center.  This mathematical convenience also somewhat simulates the observed spike in source number in the Central Molecular Zone.  The distributions of Galactocentric position (\rgal\ and $z$) and locally viewed coordinates ($\ell$ and \dsun) for the entire simulated sample are shown as the filled gray histograms in Figure~\ref{fig:vert_app}.  The vertical position distributions (bottom right panel) may be fit with a Gaussian to determine the centroid and FWHM for inter-comparison.  The histograms in Figure~\ref{fig:vert_app} represent the sum of 250 independent realizations of 40,000 simulated sources distributed randomly.  The mean and standard deviation of the fits to the vertical position distributions of the 250 independent realizations are shown in Table~\ref{table:vert_app} along with the number of objects remaining after each successive cut.

To simulate the observation function of the BGPS, we created several cuts to narrow the range of objects towards that which would be detected by the survey.  First, all sources with $|b| > 0\fdg5$ and \rgal~$< 4$~kpc were removed (resulting in the blue filled histograms in Figure~\ref{fig:vert_app}).  The latitude cut matches the BGPS survey limits, and the Galactocentric radius cut eliminates sources in the Galactic bar, whose kinematic distances are unreliable.  These criteria alone remove nearly half of the objects and narrow the vertical position distributions (see Table~\ref{table:vert_app}).  Next, Galactic longitude was restricted to the contiguous portion of the BGPS survey also observed in \hcop(3-2) with the HHT (S13), or $90\degr \geq \ell \geq 7\fdg5$.  This removes effects of incomplete BGPS coverage in the outer Galaxy and lack of distances (due to velocity information) in the innermost Galaxy.  The cumulative effect of these two cuts leaves just under one quarter of the original sources, and the distributions are shown as red hashed histograms in Figure~\ref{fig:vert_app}.

This set of sources encompasses the direct observational restrictions on the molecular layer, but the red histogram of heliocentric distance does not resemble the \dsun\ distribution of BGPS well-constrained distances, shown in Figure~\ref{fig:source_hist}(\emph{middle right}).  As was discussed in EB13, the EMAF-based prior DPDF strongly selects sources at the near kinematic distance, as distant sources are less likely to have a discernible mid-infrared contrast for comparison with the \spitzer/GLIMPSE mosaics.  Because the distribution in Figure~\ref{fig:source_hist} drops sharply for \dsun~$\gtrsim 5.5$~kpc, we modeled this by randomly removing 90\% of the sources in the red histogram in Figure~\ref{fig:vert_app} beyond \dsun~= 5.5~kpc.  This removal fraction approximates the ratio of sources per bin at 4~kpc to that at $\sim 11$~kpc in Figure~\ref{fig:source_hist}.

The final subset contains a little under 2,000 objects, comparable with the 1,710 BGPS objects with well-constrained distance estimates.  Two points bear mention.  The first is the vertical position distribution of the final subset, whose FWHM ($67.5 \pm 1.5$~pc) and centroid ($+10.9 \pm 0.8$~pc) are consistent with the measured values ($65.3 \pm 1.3$~pc and $+9.2 \pm 0.7$~pc, respectively) from \S\ref{res:vert}.  The observed narrow vertical position distribution of BGPS sources is therefore consistent with the molecular layer as probed by CO \citep[][]{Bronfman:1988} mitigated by the observational restrictions of the BGPS.  Second, in order to create an appropriately sized subset ($N\sim2000$) that meets the observational parameters of the well-constrained distance sources of Table~\ref{table:dists}, an initial sample of 40,000 dense molecular cloud structures is required.  This value represents an approximate census of the total expected number of these objects throughout the Galactic plane.

\subsection{\thco\ as Molecular Cloud Clump Tracer}\label{disc:thco}

The ubiquitous nature of CO in the Galactic plane has made it an invaluable tool for studying large-scale Galactic structure \citep[\cf][]{Dame:2001}.  The low excitation density of the optically-thinner isotopologue \thco\ explains the many velocity components identified in surveys such as the GRS along any given line of sight, especially towards the crowded inner Galaxy.  The catalog of molecular cloud clumps presented in \citet{Rathborne:2009} implies that \thco\ emission is concentrated in denser regions surrounded by an envelope of more diffuse emission, represented by the larger GMCs also cataloged by \citeauthor{Rathborne:2009}  Utilizing this emission enhancement for assigning a velocity to dust-continuum-identified molecular cloud structures has usually meant looking for the brightest emission peak along the line of sight.

Comparison between the \vlsr\ extracted directly from the \thco\ spectra of the GRS and that from a dense gas tracer (see \S\ref{data:spec}), however, yields only a $\sim 85\%$ matching rate.  This implies that while the \thco\ associated with molecular cloud clumps does produce enhanced emission, it may be outshone by expanses of less dense gas not associated with the molecular cloud clump in question, whether by quantity of gas or changes in excitation temperature due to the local environment.  Extraction of a spectrum utilizing the morphology of millimeter dust continuum emission as a prior increases the agreement rate with \hcop(3-2) to 95\%, with approximately one in eight of the disagreeing sources having a \vlsr\ outside the velocity range of the GRS data.  By identifying molecular cloud structures through millimeter dust continuum emission, it is therefore possible to convert a molecular transition line with low $n_\mathrm{eff}$ into a powerful dense gas tracer.

The morphological spectrum extraction technique is able to assign a \vlsr\ to molecular cloud structures at a high rate, exceeding 80\% for $S_\mathrm{1.1~mm} \gtrsim 0.4$~Jy, and does significantly better than dense gas tracers for $S_\mathrm{1.1~mm} \lesssim 0.1$~Jy, where the detection rate for other molecules falls below 20\%.  Especially due to the limited Galactic longitude range of the GRS, this technique does not add kinematic information for a significant number of bright BGPS sources, but it is almost exclusively the tracer of choice for low flux density objects.

While this technique was developed using the Galactic Ring Survey with its northern coverage, it is directly applicable to upcoming large CO surveys.  The Mopra Southern Galactic Plane CO Survey \citep{Burton:2013} will observe the $J=1-0$ transitions of \twco, \thco, and C$^{18}$O over the range $305\degr \leq \ell \leq 345\degr$ and $|b| \leq 0.5\degr$ with $\sim 0.7$~K rms sensitivity.  With 35\arcsec\ angular and 0.1~\kms\ spectral resolution, this survey will complement the GRS and provide a symmetric view of the molecular Galaxy into the fourth quadrant.  Additionally, the northern hemisphere JCMT/HARP \twco(3-2) survey \citep{Dempsey:2013} will cover $10\degr \leq \ell \leq 65\degr$ and $|b| \leq 0.5\degr$ with $\sim 1$~K rms sensitivity and (smoothed) 16\arcsec\ angular and 1~\kms\ spectral resolution.  The improved angular resolution of the HARP and Mopra surveys will provide CO observations more closely matched to the ATLASGAL data set.  For use with these surveys, the two tunable parameters in the morphological spectrum extraction technique (threshold $T_A$ for detection, and the primary-to-secondary peak $T_A$ ratio) will require calibration, as in Figure~\ref{fig:grs_opt}.

\subsection{Inter-Comparison of Prior DPDF Methods}\label{disc:comp}

\begin{deluxetable*}{lrrccc}
  \tablecolumns{6}
  \tablewidth{0pt}
  \tabletypesize{\footnotesize}
  \tablecaption{Application of DPDFs to BGPS V2 Sources\label{table:dpdfs}}
  \tablehead{
    \colhead{} & \colhead{} & \colhead{} & \multicolumn{3}{c}{Well-Constrained Sources Only} \\
    \cline{4-6}
    \colhead{DPDF}  & \colhead{$N$} & \colhead{$N_\mathrm{wc}$} & \colhead{Fraction of} & \colhead{Fraction of} & \colhead{Fraction of} \\
    \colhead{} & \colhead{} & \colhead{} & \colhead{Method (\%)} & \colhead{w.c.\tablenotemark{a} (\%)} & \colhead{Full V2.1 (\%)}
  }
  \startdata
  Dense Gas \vlsr & 2432 & 1254 & 51.6 & 73.3 & 14.6 \\
  \thco\ \vlsr & 958 & 338 & 35.3 & 19.8 & 3.9 \\
  \htwo\ Scale Height & 4474 & 1384 & 30.9 & 80.9 & 16.1 \\
  EMAF & 854 & 679 & 79.5 & 39.7 & 7.9 \\
  Parallax Assoc. & 292 & 291 & 99.7 & 17.0 & 3.4 \\
  \HII\ Region Assoc. & 525 & 498 & 94.9 & 29.1 & 5.8 \\
  Gemini OB1 & 49 & 49 & 100 & 2.9 & 0.6 \\
\hline
  Well Constrained & 1710 & 1710 & \nodata & 100 & 19.9
  \enddata
  \tablenotetext{a}{Fraction of the set of well-constrained sources.}
\end{deluxetable*}

\begin{deluxetable*}{lccccc}
  \tablecolumns{6}
  \tablewidth{0pt}
  \tabletypesize{\footnotesize}
  \tablecaption{Well-Constrained Fraction for Overlapping Priors\label{table:priors}}
  \tablehead{
    \colhead{} & \colhead{\htwo\ Scale Height} & \colhead{EMAF} & \colhead{Parallax Assoc.} & \colhead{\HII\ Region Assoc.} & \colhead{Gemini OB1}
  }
  \startdata
  \htwo\ Scale Height & 0.309 & 0.789 & 0.995 & 0.949 & \nodata \\
  EMAF &  & 0.795 & 1.000 & 0.855 & \nodata \\
  Parallax Assoc. &  &  & 0.997 & 1.000 & \nodata \\
  \HII\ Region Assoc. &  &  &  & 0.949 & \nodata \\
  Gemini OB1 &  &  &  &  & 1.000
  \enddata
\end{deluxetable*}

The performance of each DPDF method presented here is shown in Table~\ref{table:dpdfs}.  The first two columns of numbers describe how many objects with a DPDF from the indicated method are in the whole catalog and the well-constrained subset, respectively.  To evaluate the power of each method, the remaining columns use the number of well-constrained sources ($N_\mathrm{wc}$) as the numerator to indicate the constituent fractions of various larger sets.  The ``fraction of method'' column ($N_\mathrm{wc} / N$) indicates how well that method produces well-constrained distance estimates, while the ``fraction of w.c.'' column describes how powerful the method is for constraining distances to molecular cloud structures.  For instance, parallax association has a very high well-constrained rate (99.7\%), but its sources constitute only 17\% of the well-constrained set.  The final column in Table~\ref{table:dpdfs} specifies the fraction of the full V2.1 catalog for each method; the values are simply scaled down from the previous column.

The greatest utility of the Bayesian DPDF formalism used here is the application of multiple prior DPDFs for KDA resolution.  The use of mostly orthogonal priors allows for the maximum number of possible KDA resolutions, but there is some overlap between priors.  To illustrate the interaction of priors, Table~\ref{table:priors} computes the fraction of sources in the intersection of prior DPDFs (\ie having both) that have well-constrained distance estimates.  For clarity, only the upper portion of the symmetric matrix is shown in the table, and the diagonal elements represent the ``fraction of method'' column in Table~\ref{table:dpdfs}.  There is no intersection with the Gemini OB1 prior, as this was defined to assign a distance only to objects in that molecular cloud complex that could not be associated with one of the trigonometric parallax measurements.

The distribution of molecular gas (\htwo\ scale height) does not offer significant leverage for KDA resolution by itself, but its intended utility is limited to high-latitude objects.  For instance, of the 4,474 objects with DPDF$_\mathrm{H_2}$, some 3,094 (69\%) are in the kinematic sample, placing an upper limit on the KDA resolution rate.  Of this kinematic pool, only 1,384 (45\%) have a well-constrained distance estimate.

\begin{figure*}[!t]
        \centering
        \includegraphics[width=6.5in]{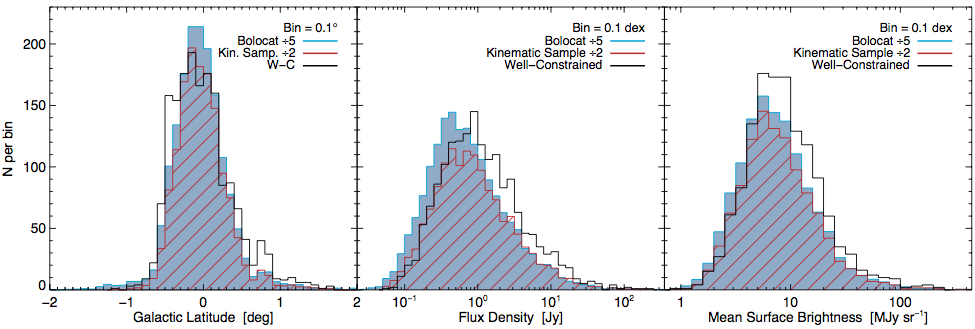}
        \caption[Comparison of observable quantities for the entire BGPS catalog, the kinematic sample, and set of well-constrained distance estimates.]{Comparison of observable quantities for the entire BGPS catalog (divided by 5, blue), the kinematic sample (divided by 2, red), and set of well-constrained distance estimates (black).  \emph{Left}: The Galactic latitude distributions.  \emph{Middle}: Source $\lambda = 1.1$~mm flux density distributions.  \emph{Right}: Source-averaged surface brightness, computed using Equation~(\ref{eqn:sb}).}
        \label{fig:catalog}
\end{figure*}

The prior DPDF based on EMAFs has a $\approx 80\%$ success rate for deriving well-constrained distance estimates, nearly identical to the rate from EB13.  The intersection of the EMAF sample with either of the catalog-based priors is small.  Maser emission and \HII\ regions are both associated with the later stages of star formation \citep[\cf][]{Battersby:2010,Dunham:2011a}, whereas EMAFs correspond to the earlier stages (EB13).  It was also noted in EB13 that bright PAH emission near $\lambda = 8$~\micron\ excited by UV photons from \HII\ regions breaks the assumption of smooth Galactic emission against which EMAFs are seen, and can skew the distance estimate returned by DPDF$_\mathrm{emaf}$.  So, while the EMAF-HRDS overlap is small, the high fraction (86\%) of well-constrained distance estimates implies that the KDA resolutions of the individual priors tend to agree, as opposing KDA resolutions would result in unconstrained posterior DPDFs.  This agreement rate is evidence that some level of localized bright 8-\micron\ emission from \HII\ regions does not significantly affect the KDA resolutions of EMAF-identified molecular cloud structures.

The new prior DPDFs introduced in this paper both show very high rates of well-constrained distance estimates.  The nearly-100\% rate for DPDF$_\mathrm{px}$ should be expected, as the precision of recent VLBI trigonometric parallaxes produces very narrow distance uncertainties.  The sole DPDF$_\mathrm{px}$ outlier is kinematically aberrant, and the DPDF formalism rightly returned an unconstrained distance estimate.  The \HII\ region prior produces well-constrained distance estimates for 95\% of the associated sources, which is slightly tempered in the EMAF-HRDS overlap, as discussed above.  The high distance-assignment rate should be viewed as an indication that the priors thusly assigned do not collide with other prior DPDFs because the coordinate-velocity association volume of \S\ref{prior:define} does not exceed the physical association scale length of HMSFRs in the Galactic plane.

\subsection{A Representative Sample?}\label{disc:rep} 

The 1,710 sources forming the BGPS V2 distance catalog provide a sizable sample for making comprehensive inferences about the Galactocentric distributions and physical properties of molecular cloud structures.  The strength of such inferences, however, hinges on how representative this sample is of the BGPS as a whole.  The relationships between the entire Bolocat, the kinematic sample of 3,508 sources, and the set of well-constrained distance estimates are shown in Figure~\ref{fig:catalog} for the distributions of Galactic latitude, flux density, and mean surface brightness.  The entire Bolocat (divided by 5) is shown in blue, the kinematic sample (divided by 2) in red, and the well-constrained distances are plotted in black.  Mean surface brightness is computed via
\beqn\label{eqn:sb}
SB = \frac{S_{1.1}}{\pi\theta_R^2} = 12.4~\mathrm{MJy}~\mathrm{sr}^{-1} \left( \frac{S_{1.1}}{1~\mathrm{Jy}}\right) \left(\frac{\theta_R}{33\arcsec} \right)^{-2} ~,
\eeqn
where $S_{1.1}$ is the BGPS integrated flux density and $\theta_R$ is the deconvolved radius of the catalog source \citep[][]{Rosolowsky:2010}.  Surface brightness is related to the source column density, but is a strictly observable quantity not requiring assumptions about the physics of the dust or environment.

The Galactic latitude distributions for the three sets of sources all have a Gaussian centroid at $b \approx -0.1\degr$, and a FWHM~$\approx 0.7\degr$,  suggesting that the prior DPDFs currently implemented do not bias our sampling of Galactic latitude.  A set of two-sample K-S tests, however, shows that the distributions are not identical at the 99.5\% confidence level or greater.  In terms of source flux density, the median values for the entire catalog, kinematic sample, and well-constrained distances are 0.62~Jy, 0.76~Jy, and 0.96~Jy, respectively.  This skew for the kinematic sample is accounted for in the kinematic detection rate as a function of flux density (Figure~\ref{fig:kin_detect}).  The further skew to higher flux density in the well-constrained sample is likely due in part to the flux-density selection effects of the EMAF prior (EB13).  For the distributions of $SB$, the differences are smaller, with median values of 6.8, 7.1, and 8.0 MJy~sr$^{-1}$ for the full catalog, kinematic sample, and well-constrained sample, respectively.  Another set of two-sample K-S tests reveal that the well-constrained sample is not drawn from the same parent population as the others at the 99.97\% confidence level or greater, but the full catalog and kinematic samples cannot be distinguished at the 95\% confidence level.  Although the flux-density distributions for these two sets are significantly different, the tendency for brighter sources to subtend larger solid angles (the Spearman rank correlation between $S_{1.1}$ and $\theta_R$ is $0.75-0.78$ for all three groups; see Equation \ref{eqn:sb}) balances this effect.  The implication is that the averaged surface brightness of sources is not a significant factor in the kinematic detection rate, but may slightly influence the ability of the currently-implemented prior DPDFs (such as DPDF$_\mathrm{emaf}$) to produce well-constrained distance estimates.

The comparison of these observable quantities suggests that the present distance catalog is biased towards brighter and denser objects than either the Bolocat as a whole or even the kinematic sample of objects with detected \vlsr.  Physical source properties derived from this catalog, therefore, are not entirely representative of the dimmer and/or more diffuse objects detected by the BGPS.  As a further check on the representative nature of this distance catalog, we compared our derived distances with those of \citet{Dunham:2011c}, who studied a subset of BGPS V1 sources using \nhhh(1,1).  Of the 456 objects presented in that paper, 169 lie within 60\arcsec\ of the peak location of a V2.1 source with a well-constrained distance estimate.  Distances derived here agree with those of \citeauthor{Dunham:2011c} at a rate of 70\%, where their KDA resolutions relied heavily upon \HI\ absorption techniques evaluated by eye.  The future development of an automated prior DPDF based on \HI\ absorption will undoubtedly resolve some of these discrepancies.


\section{SUMMARY}\label{ch3:summary}

We expanded the DPDF formalism for molecular cloud structures originally presented in EB13 by including a new kinematic distance likelihood method and additional prior DPDFs for KDA resolution.  For the specific case of the BGPS, we present an expanded distance catalog corresponding the recently-released Version 2 maps and source catalog (G13).

As a primary foundation, the DPDF formalism uses a kinematic distance likelihood within its Bayesian framework.  Molecular transition line surveys that probe dense gas (\eg \hcop(3-2); S13) have a $\sim 50\% - 85\%$ detection rate when pointed at catalog positions from continuum surveys of molecular cloud structures.  To increase the fraction of sources with kinematic information, we developed a technique for morphologically extracting a spectrum from the \thco\ data of the GRS for molecular cloud structures using the dust-continuum data as prior information.  The low effective density for excitation of \thco(1-0) requires that the emission spectrum from the more diffuse envelope around a millimeter catalog source be subtracted in order to return a single detectable \vlsr.  When compared to using the \vlsr\ of the brightest \thco\ peak along the line of sight, this morphological spectrum extraction technique increases the velocity-matching rate between \thco\ \vlsr\ and \hcop(3-2) from $\approx 85\%$ to $\approx 95\%$.  The additional velocities derived from \thco\ provide kinematic information for half again as many sources as the dense gas tracers alone, and this method is directly extensible to upcoming large CO surveys across the Galactic plane.

We also introduce a new set of prior DPDFs based on the association of molecular cloud structures with literature catalog objects with reliable distance estimates or robust KDA resolutions.  Molecular cloud structures from the BGPS are associated with these catalog entries based on a coordinate-velocity volume derived from the cumulative distributions of GMC physical properties from the GRS \citep{RomanDuval:2010} projected to the distance of the reference source.  The combined list of trigonometric parallax measurements of masers associated with HMSFRs from the BeSSeL Survey and VERA Project \citep[tabulated in][$N=103$]{Reid:2014} offers a treasure trove of gold-standard distances independent of kinematic assumptions.  The high precision of these parallax measurements translates into small heliocentric distance uncertainties.  Consequently, all but one of the 292 BGPS V2 sources associated with one or more maser parallax measurement have well-constrained distance estimates from the posterior DPDFs.  In parallel, the list of 441 \HII\ regions from the HRDS with robust KDA resolutions provides a strong constraint on distances to BGPS objects; nearly 95\% of sources lying within the association volume of one or more HRDS \HII\ region have well-constrained distance estimates.  In the outer Galaxy, the kinematic avoidance zone around $\ell = 180\degr$ prohibits the use of kinematic distances.  The BGPS-studied Gemini OB1 molecular cloud, however, lies in this region, and we assign distances to objects in this complex based on the trigonometric parallaxes to objects S252 \cite[2.10~kpc;][]{Reid:2009b} and S255 \citep[1.59~kpc;][]{Rygl:2010a}.

We present the V2 distance and velocity catalog for the BGPS, where all molecular transition-line spectroscopic observations have been aligned with the updated source catalog of G13.  The catalog and computed DPDFs are publicly available.  Of 3,508 BGPS V2 sources with kinematic information, 1,710 (49\%) now have well-constrained distance estimates.  The Galactocentric positions of these objects trace out various Galactic features, including portions of the Sagittarius, Perseus, and Outer arms, as well as the Scutum-Centarus arm between the Sun and the Galactic center.  The vertical distribution of BGPS molecular cloud structures with well-constrained distance estimates is narrower than the measured distribution of CO in the disk, but an analysis of the observational biases in the BGPS reveals that our measurement is consistent with the CO layer as measured by \citet{Bronfman:1988}.  Furthermore, in order to obtain the $\sim 2,000$ objects in the present distance catalog, there must be $\sim 40,000$ molecular cloud clumps throughout the Galactic plane.  This represents the first approximate census of the total expected number of these objects.  Finally, while the collection of well-constrained distances represents only 20\% of the BGPS catalog, the distributions of observable quantities are generally representative of the entire catalog, but are biased towards higher-density sources due to the suite of prior DPDFs currently available.

\acknowledgments

This work was supported by the National Science Foundation through NSF grant AST-1008577.  The BGPS project was supported in part by NSF grant AST-0708403, and was performed at the Caltech Submillimeter Observatory (CSO), which was supported by NSF grants AST-0540882 and AST-0838261.  The CSO was operated by Caltech under contract from the NSF.  ER is supported by a Discovery Grant from NSERC of Canada.  NJE is supported by NSF grant AST-1109116.  This publication makes use of molecular line data from the Boston University-FCRAO Galactic Ring Survey (GRS).  The GRS is a joint project of Boston University and Five College Radio Astronomy Observatory, funded by the NSF under grants AST-9800334, AST-0098562, AST-0100793, AST-0228993, \& AST-0507657.  

\appendix

\section{Glossary}\label{app:gloss}

\begin{deluxetable*}{ll}
  \tablecolumns{2}
  \tablewidth{4in}
  \tabletypesize{\small}
  \tablecaption{Glossary of Acronyms\label{table:glossary}}
  \tablehead{
	\colhead{Term} & \colhead{Definition}	
  }
  \startdata
  BeSSeL & The Bar and Spiral Structure Legacy Survey \citep[][]{Brunthaler:2011} \\
  BGPS   & The Bolocam Galactic Plane Survey \citep[][G13]{Aguirre:2011} \\
  DPDF   & distance probability density function \\
  EMAF   & eight-micron absorption feature \\
  GMC    & giant molecular cloud \\
  GRS    & The Galactic Ring Survey \citep[][]{Jackson:2006} \\
  HHT    & Heinrich Hertz Submillimeter Telescope (Mt. Graham, AZ) \\
  HISA   & \HI\ self-absorption \\
  HIE/A  & \HI\ emission / absorption \\
  HMSFR  & high-mass star forming region \\
  HRDS   & \HII\ Region Discovery Survey \citep[][]{Bania:2010,Bania:2012} \\
  IRDC   & infrared dark cloud \\
  KDA    & kinematic distance ambiguity \\
  VERA   & the Japanese VLBI Exploration of Radio Astrometry project \\
  VLBI   & very-long-baseline interferometry
  \enddata
\end{deluxetable*}
This work contains many acronyms; we present in Table~\ref{table:glossary} a glossary of terms for reference.

\section{Derivation of Kinematic Distance Likelihoods From the Reid et al. (2014) Rotation Curve}\label{app:kdist}

\begin{figure}[!t]
        \centering
        \includegraphics[width=3.1in]{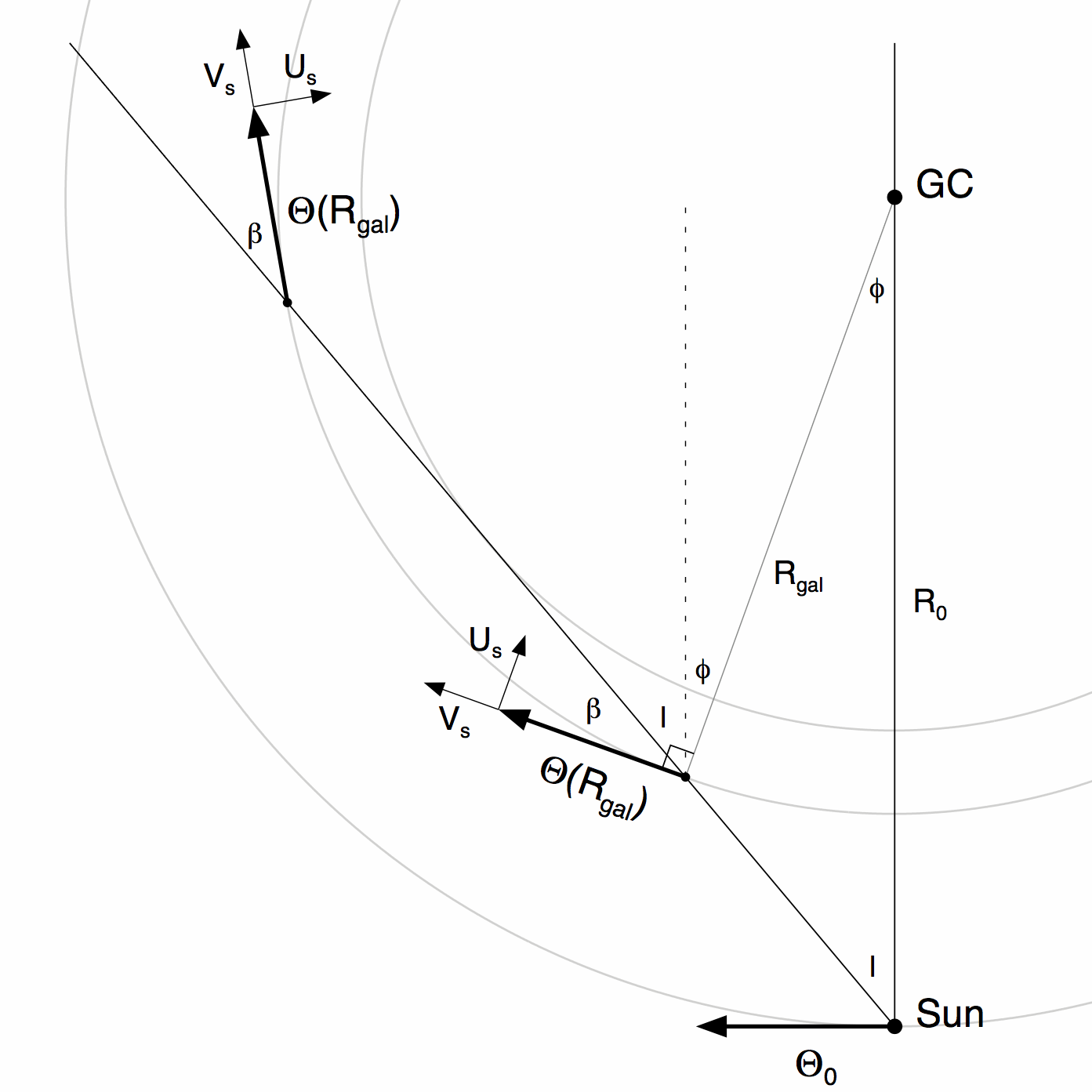}
        \caption[Galactic geometry for the computation of kinematic distances.]{Galactic geometry for the computation of kinematic distances.  Circular arcs mark orbits of constant \rgal, the outermost being the Solar circle.  Of the two inner orbits shown, the orbit with larger \rgal\ intersects the line of sight along $\ell$ twice, each with equal projected \vlsr.  The innermost orbit is tangent to the line of sight, and represents the smallest \rgal\ visible along the illustrated $\ell$.  Various velocities, angles, and distances are indicated, which are described in the text.}
        \label{fig:geometry}
\end{figure}

The rotation curve of \citet{Reid:2014} used here includes parameters describing the mean peculiar motions of HMSFRs away from the standard circular orbits about the Galactic center.  In this appendix, we derive the kinematic distance likelihood $\mathcal{L}(v_\mathrm{LSR},l,b;d_{_\sun})$ from Equation~\ref{eqn:ch3_dpdf} using the full features of the \citeauthor{Reid:2014} curve.  The relevant geometry for this derivation is shown in Figure~\ref{fig:geometry}.

The circular rotation velocity as a function of \rgal\ is given by 
\beqn
\Theta(R_\mathrm{gal}) = \Theta_0 + (d\Theta/dR_\mathrm{gal})(R_\mathrm{gal} - R_0)~,
\eeqn
where $\Theta_0$, $R_0$, and $d\Theta/dR_\mathrm{gal}$ are taken from Table~\ref{table:mw}.  By geometry, the observed line-of-sight velocity to an object along longitude $\ell$ and latitude $b$ with Galactocentric radius \rgal\ is
\begin{eqnarray}\label{eqn:app_vlsr}
v_\mathrm{obs} = (\Theta(R_\mathrm{gal}) + \overline{V_s}) \cos \beta \cos b + \nonumber \\ 
\overline{U_s} \sin \beta \cos b - \Theta_0 \sin \ell~,
\end{eqnarray}
where $\beta$ is the angle between the circular motion of the source and the line of sight, and $\overline{V_s}$ and $\overline{U_s}$ are the mean peculiar motions of HMSFRs, as discussed in \S\ref{res:rotcurve}.  Using the fact that
\beqn
\beta = \frac{\pi}{2} - \ell - \phi
\eeqn
and the trigonometric relationships for $\phi$ (the Galactocentric azimuth) are
\begin{eqnarray}
\sin \phi &=& \frac{d_{_\sun} \sin \ell}{R_\mathrm{gal}}~\\
\cos \phi &=& \frac{R_0 - d_{_\sun} \cos \ell}{R_\mathrm{gal}}~,
\end{eqnarray}
Equation~(\ref{eqn:app_vlsr}) may be written as
\begin{eqnarray}\label{eqn:app_vlsr2}
v_\mathrm{obs} = (\Theta(R_\mathrm{gal}) + \overline{V_s}) \left[ \frac{R_0}{R_\mathrm{gal}} \sin \ell \right] \cos b + \nonumber \\
\overline{U_s} \left[ \frac{R_0}{R_\mathrm{gal}} \cos \ell - \frac{d_{_\sun}}{R_\mathrm{gal}}  \right] \cos b - \Theta_0 \sin \ell~.
\end{eqnarray}
To remove \rgal\ and frame this relationship solely in terms of velocities and the distance ratio ($d_{_\sun} / R_0$), we utilize the relationship 
\beqn
R_\mathrm{gal} = \sqrt{R_0^2 + d_{_\sun}^2 - 2 R_0\, d_{_\sun} \cos \ell}
\eeqn
to recast the observed velocity as
{\small \begin{eqnarray}\label{eqn:app_vlsr3}
v_\mathrm{obs} = \left[ \frac{(\Theta(R_\mathrm{gal}) + \overline{V_s})\,\cos b}{\sqrt{1 + (d_{_\sun} / R_0)^2 - 2 (d_{_\sun} / R_0) \cos \ell}} - \Theta_0 \right] \sin \ell + \nonumber \\
\frac{\overline{U_s}\,\cos b}{\sqrt{1 + (d_{_\sun} / R_0)^2 - 2 (d_{_\sun} / R_0) \cos \ell}} \left[ \cos \ell - \frac{d_{_\sun}}{R_0} \right]~.
\end{eqnarray}}

Equation~(\ref{eqn:app_vlsr3}) describes the velocity of the source as a function of \dsun\ relative to the Solar circle orbit with velocity $\Theta_0$.  Comparison with \vlsr\ reported at the telescope from molecular transition line observations, however, requires correction for the Solar peculiar motion used by the telescope.  Rather than convert the reported velocities to match the Solar peculiar motion fit as part of the rotation curve \citep[][]{Reid:2014}, we chose to transform the velocity of Equation~(\ref{eqn:app_vlsr3}) into the (old) IAU Solar motion frame.  The proper heliocentric velocity is computed from
\beqn\label{eqn:app:vsun}
v_\mathrm{helio} = v_\mathrm{obs} - (U_\sun \cos \ell + V_\sun \sin \ell) \cos b - W_\sun \sin b~,
\eeqn
where ($U_\sun, V_\sun, W_\sun$) are the components of the Solar peculiar motion from Table~\ref{table:mw}.  Next, add back in the IAU value of 20~\kms\ in the direction of 18$^\mathrm{h}$ right ascension and +30\degr\ declination (B1950).

The correction of Equation~(\ref{eqn:app_vlsr3}) to the telescope-reported frame results in a $v_\mathrm{tel}(\ell,b$,\dsun).  We compute the kinematic distance likelihood $\mathcal{L}(v_\mathrm{LSR},l,b;d_{_\sun})$ over an array of heliocentric distance out to 20~kpc every 20~pc.  To encapsulate all information contained in the spectral line observations, we use a Gaussian fit to the spectrum, $v_\mathrm{LSR}$.  The likelihood function is then computed via
\begin{eqnarray}
&~&\mathcal{L}(v_\mathrm{LSR},l,b;d_{_\sun}) = \nonumber \\
&~&\int \exp \left[ -\frac{(v_\mathrm{tel}(\ell,b,d_{_\sun}) - v^\prime)^2}{2\sigma_\mathrm{vir}^2}   \right] v_\mathrm{LSR}~ dv^\prime~,
\end{eqnarray}
where $v^\prime$ is the dummy integration variable over the range of valid velocities, and $\sigma_\mathrm{vir}$ is the magnitude of expected virial motions within regions of massive-star formation, accounting for peculiar motions of individual molecular cloud clumps \citep[=~7~\kms;][]{Reid:2009}.

~
~
\bibliography{ms}

\end{document}